\documentclass[superscriptaddress,pra,aps,twocolumn,10pt,longbibliography]{revtex4-1}

\usepackage{amsfonts}
\usepackage{amssymb}
\usepackage{amsmath}
\usepackage{bbm}
\usepackage{graphicx}
\usepackage{verbatim}
\usepackage{microtype}
\usepackage{hyperref}
\usepackage[english]{babel}
\usepackage{natbib}
\usepackage[pdftex]{color}
\usepackage{import}
\usepackage{multirow}
\usepackage{array}
\newcolumntype{P}[1]{>{\centering\arraybackslash}p{#1}}
\usepackage{longtable}

\usepackage[braket, qm]{qcircuit}

\hypersetup{
colorlinks=true,       
linkcolor=cyan,          
citecolor=magenta,        
filecolor=magenta,      
urlcolor=cyan,           
runcolor=cyan
}

\begin{document}

\title{Hybrid quantum-classical reservoir computing for simulating chaotic systems}
\author{Filip Wudarski}
\email{fwudarski@usra.edu}
\affiliation{USRA Research Institute for Advanced Computer Science (RIACS), CA}
\author{Daniel O'Connor}
\affiliation{Standard Chartered Bank, 1 Basinghall Avenue, London, UK}
\author{Shaun Geaney}
\affiliation{Standard Chartered Bank, 1 Basinghall Avenue, London, UK}
\author{Ata Akbari Asanjan}
\author{Max Wilson}
\affiliation{USRA Research Institute for Advanced Computer Science (RIACS), CA}
\author{Elena Strbac}
\affiliation{Standard Chartered Bank, 1 Basinghall Avenue, London, UK}
\author{P. Aaron Lott}
\author{Davide Venturelli}
\email{dventurelli@usra.edu}
\affiliation{USRA Research Institute for Advanced Computer Science (RIACS), CA}

\begin{abstract}

Forecasting chaotic systems is a notably complex task, which in recent years has been approached with reasonable success using reservoir computing (RC), a recurrent network with fixed random weights (the reservoir) used to extract the spatio-temporal information of the system. This work presents a hybrid quantum reservoir-computing (HQRC) framework, which replaces the  reservoir in RC with a quantum circuit. The modular structure and measurement feedback in the circuit are used to encode the complex system dynamics in the reservoir states, from which classical learning is performed to predict future dynamics. The noiseless simulations of HQRC demonstrate valid prediction times comparable to state-of-the-art classical RC models for both the Lorenz63 and double-scroll chaotic paradigmatic systems and adhere to the attractor dynamics long after the forecasts have deviated from the ground truth. 
\end{abstract}
\maketitle
\section{introduction}

In recent years, the rapid advancement of quantum computing (QC) has attracted interest from both academia and industry, as this new computational paradigm carries the potential to transform many disciplines of science and technology. In particular, a promising avenue for utilizing QC subroutines is the data-driven field of machine learning (ML), which is currently known as quantum machine learning (QML) \cite{Biamonte_2017, Schuld_2014, Dunjko_2016, rebentrost2018quantum}. Within the field of QML, hybrid quantum-classical approaches have emerged as a viable candidate to harness the power of both classical and quantum computation effectively, especially in the Noisy-Intermediate Scale Quantum era (NISQ) \cite{Preskill_2018}, where qubit numbers are low and the decoherence and error rates are high. These approaches integrate quantum algorithms and classical machine learning techniques to improve the accuracy and efficiency of quantum model training \textit{e.g.} expectation values of multi-qubit observables (\textit{i.e.} exponentially large matrices) are directly estimated through wave function sampling, while other subroutines such as optimization can be carried out on classical devices. Furthermore, hybrid algorithms can benefit from modular and scalable structures, which can either be simulated fully on classical computers, or as a proof-of-concept on NISQ devices.

Most hybrid algorithms are a class of variational quantum algorithms (VQAs) \cite{Cerezo2020VariationalQA} that combine classical optimization routines with quantum hardware to converge towards an (ideally global) optimum of a cost function. Multiple versions of VQAs have been created, tackling various problems ranging from combinatorial optimization problems (quantum approximate optimization algorithm (QAOA) \cite{farhi2014quantum, Hadfield_2019}), to quantum chemistry (variational quantum eigensolver (VQE) \cite{Peruzzo2013AVE}). All share the property of parameterized quantum circuits that are optimized in order to find the optimum of a cost function \cite{Benedetti_2019}. 

Multiple novel machine learning algorithms have emerged that are ideal candidates to be transferred and run on quantum hardware. In particular, a non-quantum approach called reservoir computing (RC) has gained significant attention recently due to its ability to efficiently approximate complex temporal dynamics in high-dimensional data \cite{Lukoeviius2012APG, cucchi2022hands,Gauthier_2021}. RC is a paradigm that comprises of a set of recursive neural networks composed of an input layer, an intrinsic complex dynamical system known as a `reservoir' layer, and a single, trainable readout layer. The main strength of RC lies in its reservoir system, which acts as a rich temporal and spatial feature extractor, and efficiently reduces computational complexity by maintaining fixed reservoir states without the need for backpropagation or weight adjustment within the reservoir network during the training phase. Consequently, only the readout layer undergoes training, resulting in a simplified optimization process that is both computationally efficient and effective in capturing the underlying dynamics of a system. One key aspect of RC is the 
recurrent topology of the reservoir layer, which fosters the nonlinear mixing and fading memory required for effective temporal pattern recognition. Several examples of reservoir systems have been proposed in literature, including Echo State Networks (ESNs) \cite{Lukoeviius2012APG, jaeger2001echo, jaeger2007echo} and Liquid State Machines (LSMs) \cite{maass2002real}, which apply different activation functions, connectivity rules, and learning methods to achieve varying degrees of performance and robustness. 

The application of RC in multiple domains, and its architecture flexibility, has inspired researchers to provide a quantum version of the algorithm, which initially took advantage of disordered dynamics \cite{fujii2017harnessing}. Subsequently, the quantum reservoir computing framework has been extended \cite{mujal2021opportunities,mujal2023time,bravo2022quantum,pfeffer2022hybrid,pfeffer2023reducedorder, markovic2020quantum, govia2021quantum,govia2022nonlinear,nakajima2019boosting,fry2023optimizing,kubota2022quantum,suzuki2022natural}. In this paper, we aim to provide another perspective on quantum RC approaches. We propose a hybrid quantum reservoir computing (HQRC) algorithmic architecture, that has a scalable modular structure, and can easily be adjusted to accommodate hardware-efficient implementations. It relies on partial state tomography of a parameterized quantum state. Where we restrict the tomographic approach to a handful of measurements (in practice, we use $X$, $Y$ and $Z$ bases measurements on all qubits) to extract as much dynamical information as possible from quantum circuits, while keeping the protocol cost (time for circuit execution and measurements) efficient. Most importantly, the measurement outcomes are appropriately transformed to form a classical reservoir state that can carry evolution history.  

In this paper, we focus on predicting the time series behavior of chaotic systems. We choose the Lorenz63 \cite{lorenz1963deterministic} equations, one of the simplest yet-challenging paradigmatic models that serves as a low-dimensional proxy model for some features of weather dynamics, as the main test-bed for benchmarking algorithm performance. We additionally test the method on another common benchmark: the double-scroll model, to demonstrate that other chaotic systems are well-approximated by the HQRC. We extensively investigate different variations of the HQRC Ansatz, \textit{i.e.} the impact of different sets of hyperparameters, in particular, different types and number of layers inside the quantum subroutine. This analysis allows us to find setups that are operating on reasonably low-dimensional reservoirs (108 and 271 for Lorenz63 and double-scroll benchmarks, respectively) that yields forecasts that are competitive with classical methods exploiting larger reservoirs. Additionally, our study has revealed that the after hyperparameter tuning, the HQRC algorithm provides accurate long-term reproduction of attractors. Finally, we run a proof-of-concept experiment on {\it Lucy} chip provided by Oxford Quantum Circuits. 

The paper is organized as follows: First, we provide an introduction to standard reservoir computing algorithms, delving into their underlying principles and construction. Subsequently, we shift our attention towards the hybrid quantum model, where we elucidate the integration of quantum principles with classical reservoir computing methods to create quantum-enhanced learning models. Following this, we present a set of simulation results that shed light on the performance, scalability, and robustness of the proposed hybrid model in diverse learning scenarios. These proof-of-concept results offer valuable insights into the efficacy of our quantum-classical approach but also serve as a benchmark to gauge its potential capabilities in comparison with existing classical RC techniques. Finally, we provide concluding remarks, where we outline the implications of our findings for future research, improvement and deployment of our methods.

\section{(Classical) reservoir computing}
Reservoir Computing (RC), a paradigm within Recurrent Neural Networks (RNNs), has emerged as a powerful and efficient computational approach designed for the modeling and prediction of complex, time-dependent data sequences \cite{schrauwen2007overview, verstraeten2007experimental}. RC is an umbrella category of recurrent models including approaches such as Echo State Networks (ESN) \cite{jaeger2001echo} and Liquid State Machines (LSM) \cite{maass2002real}. The primary strength of RC is the fixed, randomly initialized components called the `input' and `reservoir' layers, which provide a complex forward graph capable of encoding temporal dependencies and feature correlations \cite{pathak2018model}. Due to the random and unconventional connections in input and reservoir layers, it is impractical to calculate gradients and perform standard backpropagation, making these layers untrainable \cite{freiberger2020training}. Given an input vector $X_t$, the reservoir layer maintains the temporal dynamic in the recursively generated state $r_t$ by combining the previous state vector $r_{t-\Delta t}$ and a non-linear function of the input vector and the previous recurrent state
\begin{equation}\label{eq:reservoir_classic_state}
    r_t = (1-\alpha) r_{t-\Delta t} + \alpha f\left(W_r r_{t-\Delta t} + W_X X_t \right)
\end{equation}
where $W_X$ and $W_r$ represent the input and reservoir layers, respectively and $\alpha\in(0,1]$ is referred to as the leak rate, governing the rate of new information leakage into the system. In the operational setup, one discretizes the modeled evolution and creates the reservoir state vector for each discrete time steps, i.e. $r_{t-\Delta t}$ becomes $r_{t-1}$, where time takes integer values $t=1,2,\ldots$. 
The states of the model are calculated autoregressively and can then be mapped to output data via an output layer called `readout'. The readout layer is a ridge regression layer mapping the state vectors 
\begin{equation}
    R_t = (1, r_t, X_t)^T,
\end{equation}
to the output data ($y_t$), where the unit element is playing the role of bias. The readout layer defines a linear set of equations for state-output mapping and it is often combined with Tikhonov regularization (widely known as ridge regression) to generalize the performance and prevent overfitting
\begin{equation}\label{eq:reservoir_classic_readout1}
\hat{W}_o = \arg \min_{W_o} \left\{ \lVert \mathbf{y} - W_o\mathbf{R}] \rVert_2^2 + \beta \lVert W_o \rVert_2^2 \right\}. \\
\end{equation}
where
\begin{eqnarray}
    \mathbf{R} & = & [R_1, R_2, \ldots, R_n],\\
    \mathbf{y} & = & [y_1, y_2, \ldots, y_n]
\end{eqnarray}
are matrices of $R_t$ and $y_t$ vectors, respectively, arranged sequentially for the training part.
Where the $W_o$ and $\hat{W}_o$ represent the readout parameter and the optimized version. The $y$, $R$ and $\beta$ are the target (training) vectors, state vectors (arranged in a matrix with columns corresponding to different time instances $t$ of $y_t$ and $R_t$), and regularization parameter, respectively. Equation \eqref{eq:reservoir_classic_readout1} yields the following vectorized solution
\begin{equation}\label{eq:reservoir_classic_readout2}
        \hat{W}_o = \mathbf{y}\mathbf{R}^T (\mathbf{R}\mathbf{R}^T + \beta I)^{-1}\,,
\end{equation}
where $I$ is the identity matrix. The combination of the steps laid out in equations \eqref{eq:reservoir_classic_state} and \eqref{eq:reservoir_classic_readout2} provides the means for RC to learn and infer dynamical evolution.

\section{Hybrid quantum reservoir computing}
\begin{figure*}
    \centering
    \includegraphics[width = 0.98\textwidth]{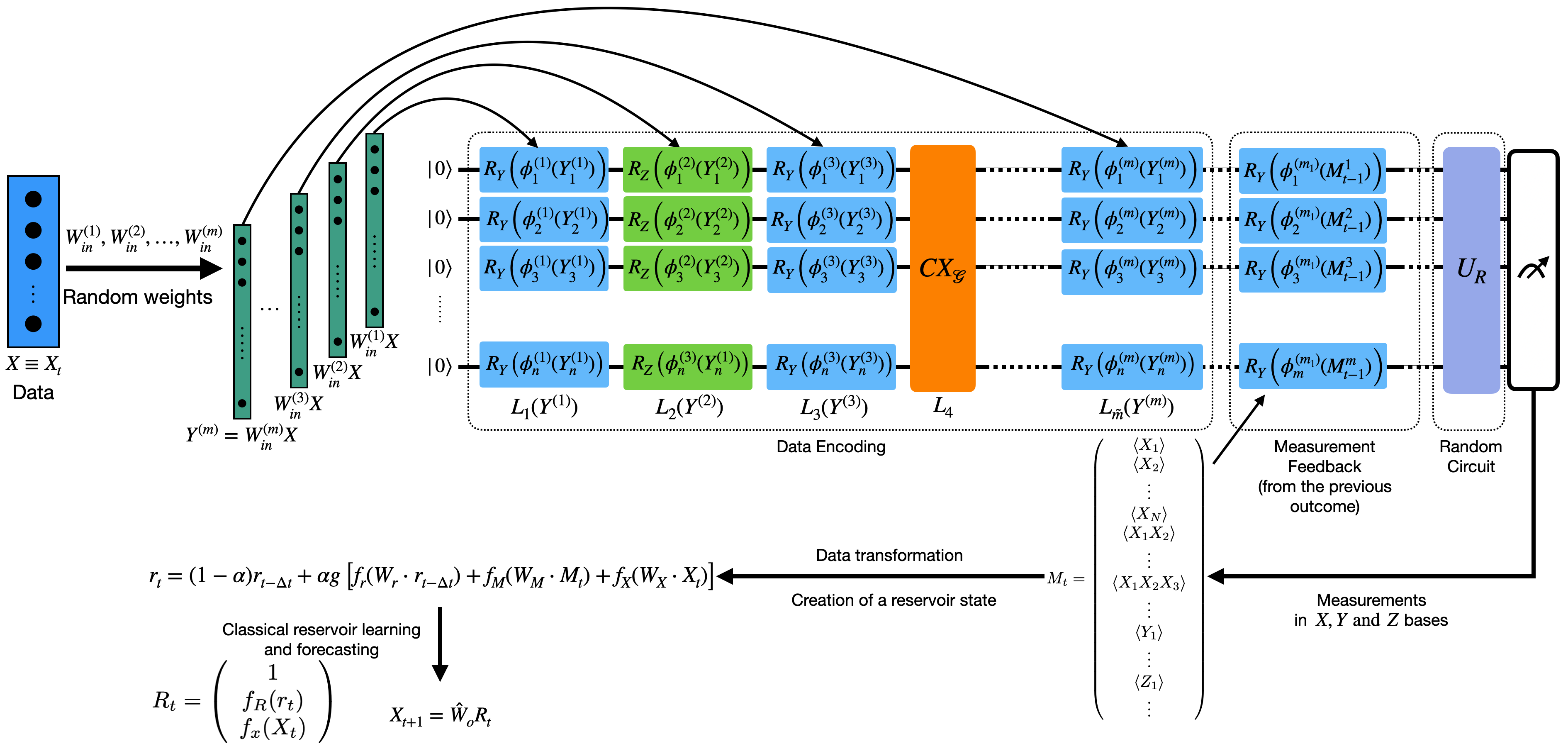}
    \caption{A schematic chart of HQRC approach. The data $X_t$ at time $t$ is transformed by fixed weight matrices $W_{in}^{(j)}$ to create vectors of data encoded parameters $Y^{(j)}$, which are distributed across the quantum circuit in parameterized layers $L_j(Y^{(j)})$ (in the chart we depict single-qubit rotations around $Y$ and $Z$-axes). The circuit also possesses parameter-free layers ($L_4$) of $CX$ (CNOT) gates, acting on qubits given by a graph $\mathcal{G}$. Subsequently, the circuit has layers related to measurement feedback that take outcomes from the previous time step ($t-1$) and encodes them in parameterized gates (here depicted as taking $M_{t-1}$ vector values, but it can also use $r_{t-1}$ as parameters). In both data encoding and measurement feedback layers, one can additionally transform the parameters with feature maps, given by functions $\phi_k^{(j)}$. Finally, a random reservoir layer is applied that takes the form of a network of random gates (\textit{e.g.} single-qubit rotations with random angles followed by a network of CX gates). The circuit is measured in fixed, specified bases (we usually choose $X, Y$ and $Z$ bases) in order to create a measurement vector, that comprises single-qubit expectation values and multi-qubit correlators defined on a measurement graph, \textit{e.g.}, $\langle X_i\rangle,\, \langle X_i X_j\rangle,\, \langle X_i X_j X_k\rangle$. The outcomes are then combined classically to generate the next reservoir state Eq.~\eqref{eq:reservoir_qrc_state}, and are used in the ridge regression procedure to determine $\hat{W}_o$ for future predictions.}
    \label{fig:qrc_chart}
\end{figure*}

In this section, we provide a full description of the proposed HQRC algorithm. The aim of the method is to predict next steps in the evolution of a dynamical system. In particular, for a time-dependent system $X_t$, we use a fixed number of observations $n$, at times: $t_1, t_2, \ldots, t_n$ to learn intrinsic relationships in the observed data to obtain new data points for $t_{n+1}, t_{n+2}, \ldots, t_{n+p}$, that ideally match the true dynamics of the system. The algorithm is composed of intertwined classical and quantum subroutines (see Fig.~\ref{fig:qrc_chart}). In the first step, one needs to transform data $X_t$ into a format that can fit the quantum subroutine. The quantum circuit comprises multiple layers that have fixed parameterized structure, though the parameters can (and in reality do) vary between each time step. The layers can be divided into three main categories: i) data encoding, ii) measurement feedback, and iii) random circuit layers. Data encoding layers can be either parameterized or parameter-free. For the former, we use linearly transformed data as input at time $t$, \textit{i.e.} $Y_t^{(j)}= W_{in}^{(j)}X_t$, where $W_{in}^{(j)}$ is a fixed matrix (usually, but not always, taken to be random as in the case of classical RC) of size $d_{L_j}\times d_{\mathrm{input}}$, which may be additionally transformed by a feature map $\phi$ (see Appendix~\ref{app:layers} for more details on tested layers). Where $d_{L_j}$ being the number of parameters that the data encoding layer $L_j$ can accommodate (\textit{e.g.} for a system of $n$ qubits, a layer of single qubit $X$-rotations, will have $d_{L_j}=n$), and $d_{\mathrm{input}}$ is the dimension of the input vector, \textit{i.e.}, $X_t$. To avoid large numerical values, we normalize $W_{in}$ matrices by their largest singular values. In our experiments, we restrict ourselves to layers composed of parameterized $R_X$, $R_Y$ or $R_Z$ rotations and parameter-free $CX$ gates, though this set can be expanded to arbitrary gates, in particular to a set which matches native gates available on a given NISQ device. In the parameter-free case, one needs to additionally define a graph of qubits upon which the gates act. In general, the graph should also be defined for parameterized gates as well. However, since we use single-qubit rotations, we assume that they act on every qubit in the quantum register. The graph can be selected to match the hardware-specific topology and therefore avoid quantum operations such as SWAP gates. Layers of $CX$ are crucial for introducing entanglement in the circuit, enabling exploration of a larger portion of the Hilbert space. In principle, this gives higher expressibility of the Ansatz.

A similar high-level structure is present in layers ii) and iii). However, instead of taking transformed $X_t$ as their parameters (if the layers utilize parameters), they rely on the measurement induced parameters for type ii) layer and randomly initialized gates for iii) layers. In the case of measurement feedback layers, one can feed the previous iteration measurement vector $M_{t-1}$ or previous reservoir state $r_{t-1}$.  The sizes of either $M_{t-1}$ or $r_{t-1}$ can differ from the number of parameters in that layer, the practitioner can select which and how many of the components are ultimately utilized (\textit{e.g.} in some experiments we use only single-qubit expectation values \textit{e.g.} $\langle X_i\rangle$ to supplement single-qubit rotation on $i$-th qubit, \textit{i.e.} $R_X(\langle X_i\rangle)$). Furthermore, all parameterized layers that are not fixed, come with extra parameter transformation, that follows the feature map encoding $\phi$ approach \cite{Mitarai_2018, vidal2020input,Kyriienko_2021,Schuld_2019} (potentially different across layers or qubits). These additional transformations can introduce non-linearities that are essential for learning complex dynamics. 

The prepared quantum circuit is measured according to a selected scheme that defines the reservoir size in the HQRC. In order to optimize measurement efficiency while extracting sufficient amount of information from the circuit, we restrict ourselves to measuring all qubits in $X, Y$ and $Z$ bases, although other bases are viable candidates, however, one needs to bare in mind that more sophisticated measurement schemes can be more resource consuming (e.g. full state tomography has been employed in \cite{pfeffer2022hybrid,pfeffer2023reducedorder}). This allows us to form a measurement vector, that combines single-qubit expectation values, \textit{i.e.} $\langle X_1\rangle, \langle X_2\rangle,\ldots, \langle X_n\rangle, \langle Y_1\rangle, \ldots, \langle Y_n\rangle, \langle Z_1\rangle,\ldots, \langle Z_n\rangle$, and multi-qubit correlators featuring the same type of Pauli matrices. The latter operates on user-defined {\it connectivity} graphs, where vertices are qubits, and edges determine the correlator. For example, for a graph $\mathcal{G}=\{ (1,2), (2,3)\}$ one extracts correlators $\langle X_1 X_2\rangle,\langle X_2X_3\rangle$, similarly for $Y$ and $Z$ operators. Note, that each expectation value, in principle, can be expressed as a Fourier transform \cite{Schuld_2021} with different form. This, ultimately, translates into having the measurement vector composed of non-linearly transformed inputs. Since, we use not only feature map encoding, but also a linear transform $W_{in}$, we avoid trivial Fourier frequencies, that can be easily simulated on a classical computer at scale \cite{xiong2023fundamental}.  Therefore, this approach provides many options for the choice of measurement scheme. In our experiments we restricted analysis up to three-body correlators defined on all-to-all connected graphs, which yields a reservoir size of at most $3N + 3{N\choose 2} + 3{N\choose 3}$. Note, that all these correletors can be efficiently calculated from just three measurement bases on the hardware. 

The measurement vector is then used to create the reservoir state, denoted as $r_t$, as follows
\begin{align}
    r_t &= (1-\alpha) r_{t-\Delta t} + \alpha g\Big[f_r(W_r \cdot r_{t-\Delta t})\nonumber\\ 
    &+ f_M(W_M\cdot M_t) + f_X(W_X\cdot X_t)\Big]\,. \label{eq:reservoir_qrc_state}
\end{align}
Here, $\alpha\in[0,1]$ represents the leak rate, a parameter that controls the memory retention within the reservoir. The functions $f_r$, $f_M$, and $f_X$ denote activation functions, which can introduce non-linearities into the system dynamics. The matrices $W_r, W_M$, and $W_X$ correspond to fixed (usually random or identity) matrices associated with the reservoir state $r_t$, the measurement vector $M_t$, that implicitly depends on $X_t$ (and previous outcomes, if measurement feedback layer is present), and the input state $X_t$, respectively. Additionally, we allow a {\it global} (non-linear if desired) transformation of the current contribution, by a function $g(\cdot)$. This equation encapsulates the critical step where quantum measurements are integrated into the reservoir state, infusing it with valuable information extracted from the quantum domain while allowing for the incorporation of non-linear transformations in the process. It is worth stressing that Eq.~\eqref{eq:reservoir_qrc_state} gives us sufficient generalization to tune the contribution from each component of that equation such that one can restore the classical reservoir computing Eq.~\eqref{eq:reservoir_classic_state}, if $f_M \equiv 0$ and $f_r=f_X=id$. Additionally, taking inspiration from classical reservoir computing, we renormalize the weight matrices by their largest singular value. This procedure, even though unnecessary in all cases, stabilizes the method, as the training and prediction steps stay within reasonable boundaries.

The final stages of our classical processing are characterized by operations that are efficient to simulate. At this point, the obtained reservoir states, in conjunction with the input states, are combined to form a vector $R_t$ that plays a pivotal role in the subsequent learning procedure
\begin{equation}\label{eq:learning_vect_qrc}
    R_t = \left(1, f_R(r_t), h_X(X_t) \right)^T\ .
\end{equation}
This learning procedure is governed by the ridge regression method, a well-established technique in machine learning. It is notable the potential for introducing additional non-linear transformations to the vector before it is utilized in ridge regression in the form of $f_R$ and $h_X$. The unit value prepended to the $R_t$ vector takes the role of a bias. This affords our model a unique degree of flexibility, enabling it to capture and exploit complex, non-linear relationships within the data, thereby enhancing its capacity for accurate and nuanced learning. Similarly, though focusing on slightly different exploitation of non-linearity, work \cite{Gauthier_2021} demonstrated improvements in learning capabilities for classical approaches, if second-order contribution (\textit{i.e.} squared reservoir states) is taken into account.

The introduced method places different stress on the meaning of hybrid quantum algorithms. As it takes inspiration from other techniques such as QAOA, VQE, or Quantum Neural Networks (QNN) \cite{Nguyen2022TheoryFE,Abel2022CompletelyQN}, it additionally relies more heavily on the incorporation of quantum features from measurements. Instead of focusing on extracting ``physics-inspired'' information, that is related to a problem to be solved, \textit{e.g.} Hamiltonians in VQEs, it takes various measurements as proxies carrying potentially relevant information. Note that we treat the set of measurement operators as an additional hyperparameter that is arbitrarily chosen, instead of deliberately selected based on the problem. The measurements enable us to combine classical reservoir computing principles with the inherent computational capabilities of quantum systems. The core innovation of our method lies in its ability to map intricate problem spaces onto a high-dimensional Hilbert space constructed within a quantum circuit. It is still an open question, if the HQRC networks at scale suffer from concentration phenomenon \cite{xiong2023fundamental}.

Measurement plays a  pivotal role in our method: measurements are efficiently computable within the quantum framework, but they also are introducing non-linearities into the system (each component is a unique Fourier series). This introduction of non-linearity enables us to capture and manipulate complex relationships within the data, which may not be amenable to linear transformations. Furthermore, our method offers the flexibility to further transform these measurements as needed, allowing us to tailor the analysis to the specific characteristics of the problem at hand. This adaptability in measurement transformation ensures that our method can effectively address a wide spectrum of machine learning challenges, from linear to highly non-linear, whilst also maintaining computational efficiency and interpretability.

\section{Results}
In this section, we present the results of the HQRC approach applied to chaotic systems. We use the Lorenz63 benchmark as the main test-bed for the algorithm. However, we test our method also on other chaotic systems double-scroll, which is also a three-dimensional problem. Our analysis encapsulates simulations of quantum circuits on a classical computer, where we investigate various setups of hyperparameters. Finally, we show proof-of-concept results collected from 8 qubit {\it Lucy} chip, provided by Oxford Quantum Computing. 

\subsection{Metrics}
As the main metric that describes quality of the network, we use the valid prediction time (VPT) \cite{vlachas2020backpropagation, platt2022systematic}. The VPT is a time instance $t$, when the deviation between simulated predictions and the ground truth exceed a set threshold with respect to the root mean square error (RMSE)
\begin{equation}
    RMSE(t) = \sqrt{\frac{1}{D}\sum_{i=1}^D\left(\frac{\tilde{y}_i(t) - y_i(t)}{\sigma_i} \right)^2}\ge \varepsilon,
\end{equation}
where $\tilde{y}_i(t), y_i(t)$ are $i$-th components at time $t$ of predictions and the ground truth, respectively, $\sigma_i$ is the $i$-th component of standard deviation of the true data serving as normalization and $D$ is the dimensionality of the problem (\textit{e.g.} for Lorenz63 and double-scroll $D=3$). In our analysis, we select $\varepsilon = 0.3$, following a systematic review of classical reservoir computing in \cite{platt2022systematic}.  

In the case of chaotic systems, it is clear that one cannot expect indefinite forecasting. Therefore, an equally important metric for benchmarking these systems is long-term attractor prediction, which means that the system stays in its basins of attractions, while potentially deviating from the correct component-wise predictions. Hence, in our analysis, we investigate the closeness of predicted and ground truth attractors. Additionally, we use Poincar\'e return map \cite{Pathak_2017} to order all local maxima of the predicted and actual time series (for that we use longer simulations) and order them as $[z_1,z_2,\ldots, z_m]$ (for $z$ component of the Lorenz63 vector, where the subscript denotes $m$-th maximum of the prediction phase) and plot them against each other $[z_i, z_{i+1}]$.

\subsection{Classical simulations}
Since the HQRC provides sufficient flexibility in defining the number of qubits, depth of the circuits as well as the number of components in the measurement vector, which ultimately translate into the size of the reservoir, one may select them such that they can be simulated on a classical computer. In this section, we analyze results that demonstrate that the given framework is capable of providing sufficient expressibility to predict behavior of chaotic systems that are comparable with the state-of-the-art results with classical reservoir computing. 

\subsubsection{Lorenz63}\label{sec:classical_L63}
Lorenz63 \cite{lorenz1963deterministic} is a standard benchmark for classical RC, as it is well-studied chaotic model. The dynamics of the system is governed by the following set of differential equations
\begin{eqnarray}
    \frac{d x(t)}{d t} & = & 10 \left[y(t)-x(t)\right],\nonumber\\
    \frac{d y(t)}{dt} & = & x(t)\left[28-z(t) \right] - y(t),\label{eq:L63}\\
    \frac{d z(t)}{dt} & = & x(t)y(t)-\frac{8z(t)}{3},\nonumber
\end{eqnarray}
where we fixed coefficients to match the commonly used values in the literature \cite{Gauthier_2021}. These equations are analogous to a simplified weather model of atmospheric convection that experience uniform heating and cooling from below and above, respectively. In our experiments we use the following initial conditions $x(0)=17.67715816276679$, $ y(0)=12.931379185960404$, and  $z(0)=43.91404334248268$ (we have tested other (randomly chosen) initial conditions, which yield similar VPT values). The Lorenz63 vector is normalized to be confined to [-1,+1] region during the training stage, which is achieved by diving all values by the largest absolute value among all components in the training set. 

\begin{figure*}[t]
    \centering
    \includegraphics[width = 0.98\textwidth]{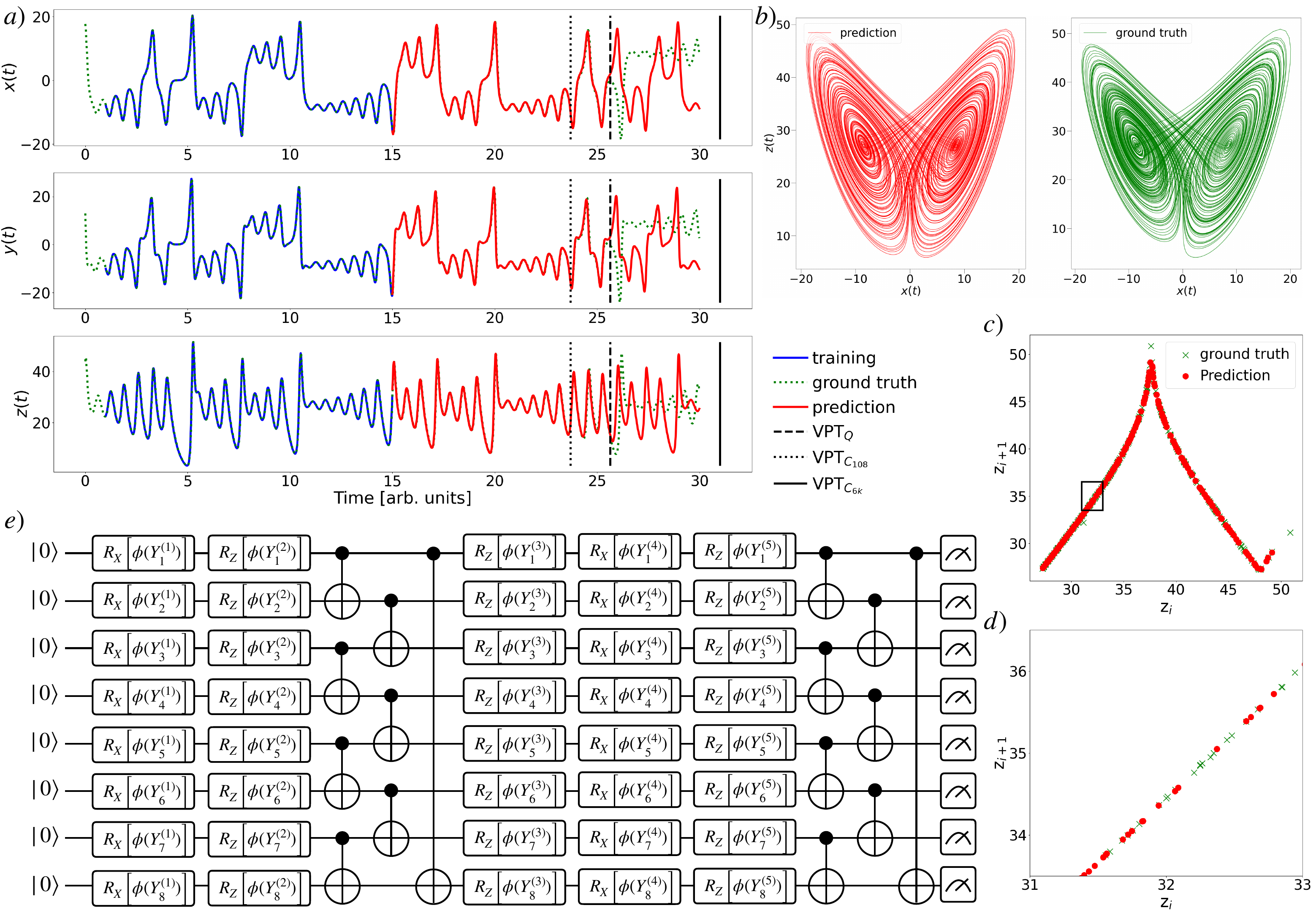}
    \caption{Simulated results of Lorenz63 chaotic system of reservoir computing.  a) the HQRC approach simulated with circuit structure depicted in e), where $Y_k^{(j)}= W^{(j)}_{in}X$ is a data encoding vector, transformed with feature function $\phi(x) = \tanh(x)$. We use 1500 training points (with time increment of $dt=0.01$ in arbitrary units) and 1500 time steps for the prediction phase. The Long-term qualitative correct behavior reconstruction is depicted in panel b) for attractor position, as well as for Poincar\'e return map in panel c) and zoomed-in version d). This indicates that even component-wise divergence after around 25 time units, the chaotic evolution displays appropriate oscillatory return pattern. We compare the results with VPT values for reservoir size of 6000 from \cite{platt2022systematic} (solid line $C_{6k}$) and for our own RC for reservoir states of size 108 (dashed line $C_{108}$). The circuit is measured in $X,Y$ and $Z$ bases in order to extract single-qubit expectation values, and two-qubit correlators for the fully connected graph.}
    \label{fig:best_result}
\end{figure*}

In Fig.~\ref{fig:best_result} we present behavior of a well-performing setup for the Lorenz63 problem, that we have identified in this study (see Appendices for more setups). The setup is composed of 8-qubits, and the quantum circuit is composed solely of data encoding layers, which means that the recurrence only takes place on the classical processing side. We have identified (see Appendix~\ref{app:layers}) that, in case of Lorenz63 we obtain statistically better results for reservoir Ans\"atze without the measurement feedback layers, while the presence of random circuit part can improve VPT values at the expanse of deeper circuits (see Appendix~\ref{app:layers} for statistical analysis). The outcomes are collected from a noiseless simulator without shot-noise, which means, the expectation values have been calculated from the wave function, instead of sampled with a finite number of shots. In Fig.~\ref{fig:best_result} we compile: a) short-time predictions of all three components of Lorenz63 vector, and long-term predictions in panels b) for attractor reconstruction and c)-d)  Poincar\'e return map. The system comprises of total 7 data encoding layers (panel e)), from which there are 2 parameter-free layers of CX network with underlying graph having connectivity of a ring. Additionally, the remaining 5 layers are single qubit rotations (see Appendix~\ref{app:hyperparams} for further details about hyperparameters). In that setup, since we use $X, Y, Z$ measurements to construct the measurement vector composed of single qubit expectation values and two-qubit correlators between qubits from the fully connected graph, we utilize reservoirs of 108 size only. Despite the small size, for the best hyperparameters, we obtain VPT of 10.68, which is better than the state-of-the-art classical approaches with comparable reservoir sizes (see for comparison \cite{platt2022systematic, Pathak_2017}, in the plot we included our RC simulations that yield better results than one reported in \cite{platt2022systematic} for reservoir's size 108). However, if larger reservoirs are utilized for classical systems, we observe that it is possible to obtain higher VPT values of $\approx$16. Note, that the reservoir size is not the only important property of these algorithms. However, it is the reservoir size that primarily determines the cost of the most computationally demanding subroutine that scales as $O(Mn^2)$, where $n$ is the reservoir size and $M$ is the number of training steps, which is linked to matrix inversion in the ridge regression \cite{Gauthier_2021}. Moreover, if taken the same HQRC hyperparameter setup with $f_M(W_M\cdot M_t) \equiv 0$ for each $t$, i.e. to reconstruct the {\it classical limit}, we obtain values substantially deviating from the ground truth with VPT$=0$.

\begin{figure}
    \centering
    \includegraphics[width = 0.48\textwidth]{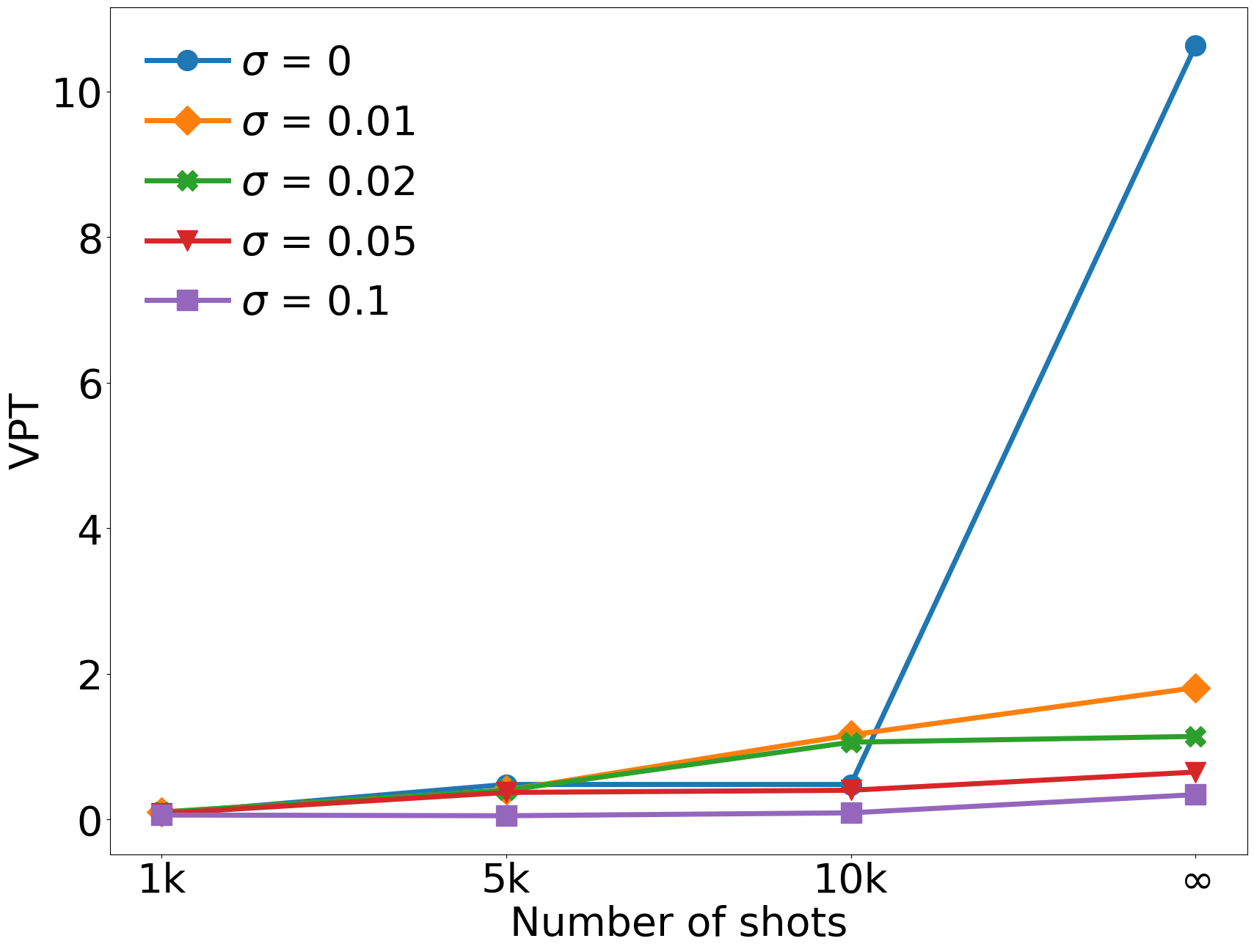}
    \caption{Performance of the HQRC algorithm with quantum circuit given in Fig.~\ref{fig:best_result} e) for varying number of shots ($\infty$ indicates the exact expectation values), for noiseless case $\sigma=0$ and simulation with extra Gaussian coherent noise in rotation angles with standard deviation $\sigma$ (see Appendix~\ref{app:hyperparams} for further analysis on the number of shots).}
    \label{fig:shots_scaling}
\end{figure}

Recognizing that having noiseless simulation with exact expectation values is an idealized scenario, we investigate performance of the algorithm with finite samples, and a certain amount of coherent noise in encoding layers. We inject mispecification noise, as Gaussian noise random variable (leading to over- or under-rotations) centered around the ideal value with standard deviation $\sigma$. In Fig.~\ref{fig:shots_scaling} we see that restricting to a reasonably large number of shots (10,000) the performance already decreases significantly. Similar detrimental effects are recorded for coherent noise, even when standard deviation is small. The limited number of shots, not only affect accuracy of the expectation values (effectively making results statistically unstable), but also forces them to be determined with finite decimal precision. The latter is crucial in case of predicting chaotic dynamics, as small deviations in initial conditions (or as in that case, throughout the training process) can lead to higher discrepancies in trajectories. Additionally, it is an open problem to identify how many samples one needs to collect to obtain decent results, and if the number of samples scale polynomially with the number of qubits in the HQRC network, otherwise concentration phenomena can stifle the performance \cite{xiong2023fundamental}.

The presented results provide preliminary evidence to support that the HQRC approach is a viable quantum alternative to the classical RC approaches. In particular, the standard RC approach is sensitive to various hyperparameters (reservoir size, training length, etc. \cite{platt2022systematic}). As the proposed method has multiple hyperparameters to select (number and type of layers, type of feature maps, measurement correlators, etc.), we observe a hyperparameter sensitivity as well, however even though the variations between performance can fluctuate, the forecasts rarely diverge from stable solution, which is not always the case for the classical RC. 

\subsubsection{Double-scroll}
Another popular benchmark is based on the dynamics of a double-scroll electronic circuit given by 
\begin{eqnarray}
    \frac{dV_1(t)}{dt} & = & \frac{V_1(t)}{R_1} - \frac{\Delta V(t)}{R_2} - 2I_r \sinh\left(\beta\Delta V(t)\right),\nonumber\\
    \frac{dV_2(t)}{dt} & = & \frac{\Delta V(t)}{R_2} + 2 I_r \sinh\left(\beta \Delta V(t)\right)- I(t),\label{eq:DS}\\
    \frac{d I(t)}{dt} & = & V_2(t) - R_4 I(t)\nonumber,
\end{eqnarray}
in dimensionless form, with $\Delta V(t) = V_1(t)-V_2(t)$. We fixed the parameters to: $R_1=1.2, R_2 = 3.44, R_4 = 0.193,\beta = 11.6$ and $I_r=2.25\times10^{-5}$, and we discretize evolution into $dt=0.25$ increments following \cite{Gauthier_2021}, and initial conditions as $V_1(0)=0.37926545$, $V_2(0)=0.058339$, $I(0)=-0.08167691$.

\begin{figure}[t]
    \centering
    \includegraphics[width = 0.49\textwidth]{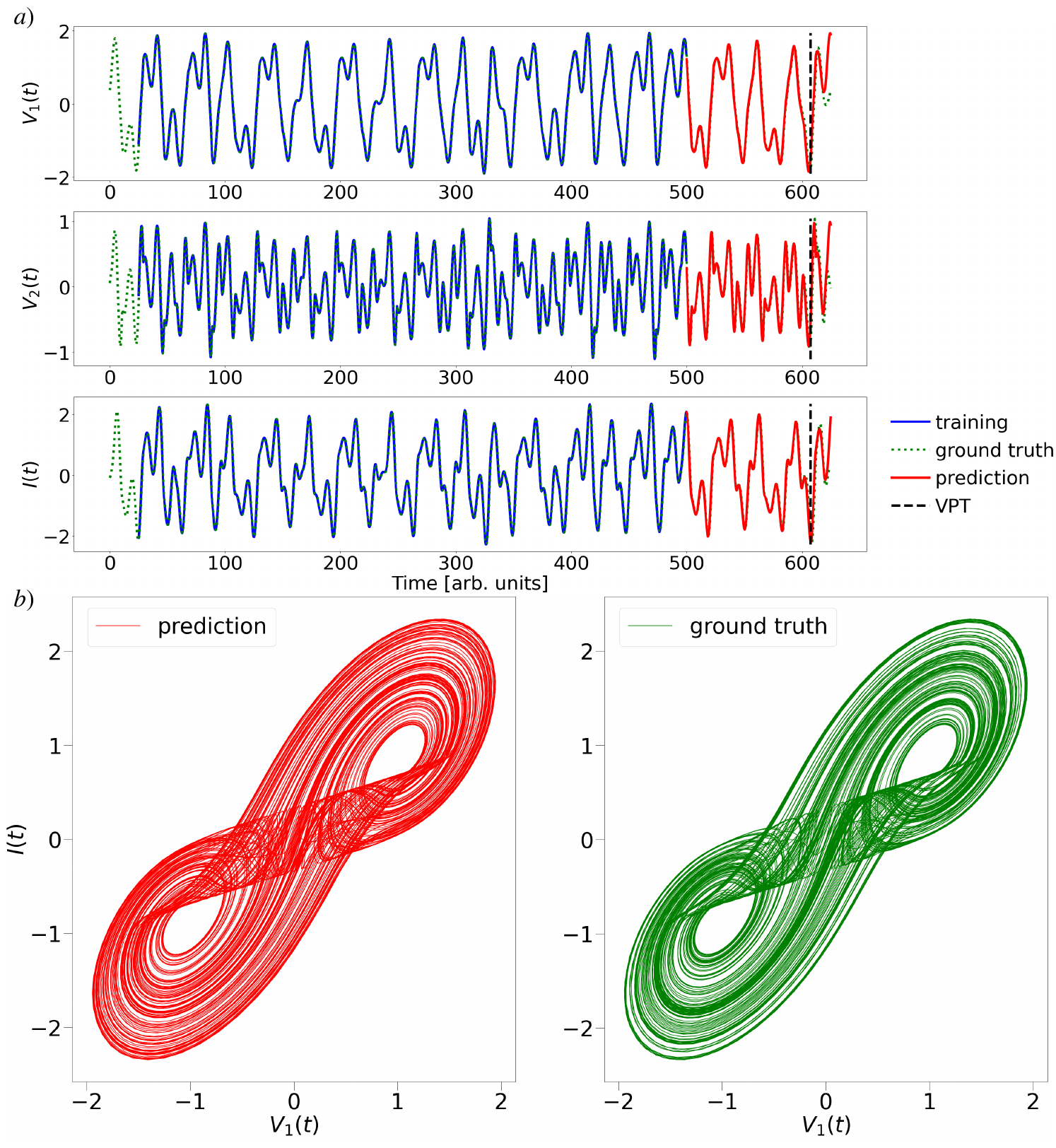}
    \caption{Simulated results for double-scroll system with 8 qubit. In panel a) we see training that overlaps with the ground truth and prediction steps that start to deviate after around 100 time units (VPT is 107.5). In the panel b) we depict reconstruction of long-term behavior based on the attractor.}
    \label{fig:ds_best}
\end{figure}

Inspired by the Ansatz incarnation for the Lorenz63 model (see Appendix~\ref{app:layers}), we performed a restricted search for well-performing hyperparamters. In Fig.~\ref{fig:ds_best} we show performance of the HQRC algorithm with 8 qubits and reservoir size of 271. This leads to VPT value of 107.25, that is also competitive with state-of-the-art results \cite{Gauthier_2021}.

Further details on the hyperparameters and different setups are available in Appendix~\ref{app:hyperparams}. 

\subsection{QPU results}
\begin{figure*}
    \includegraphics[width = 0.98\textwidth]{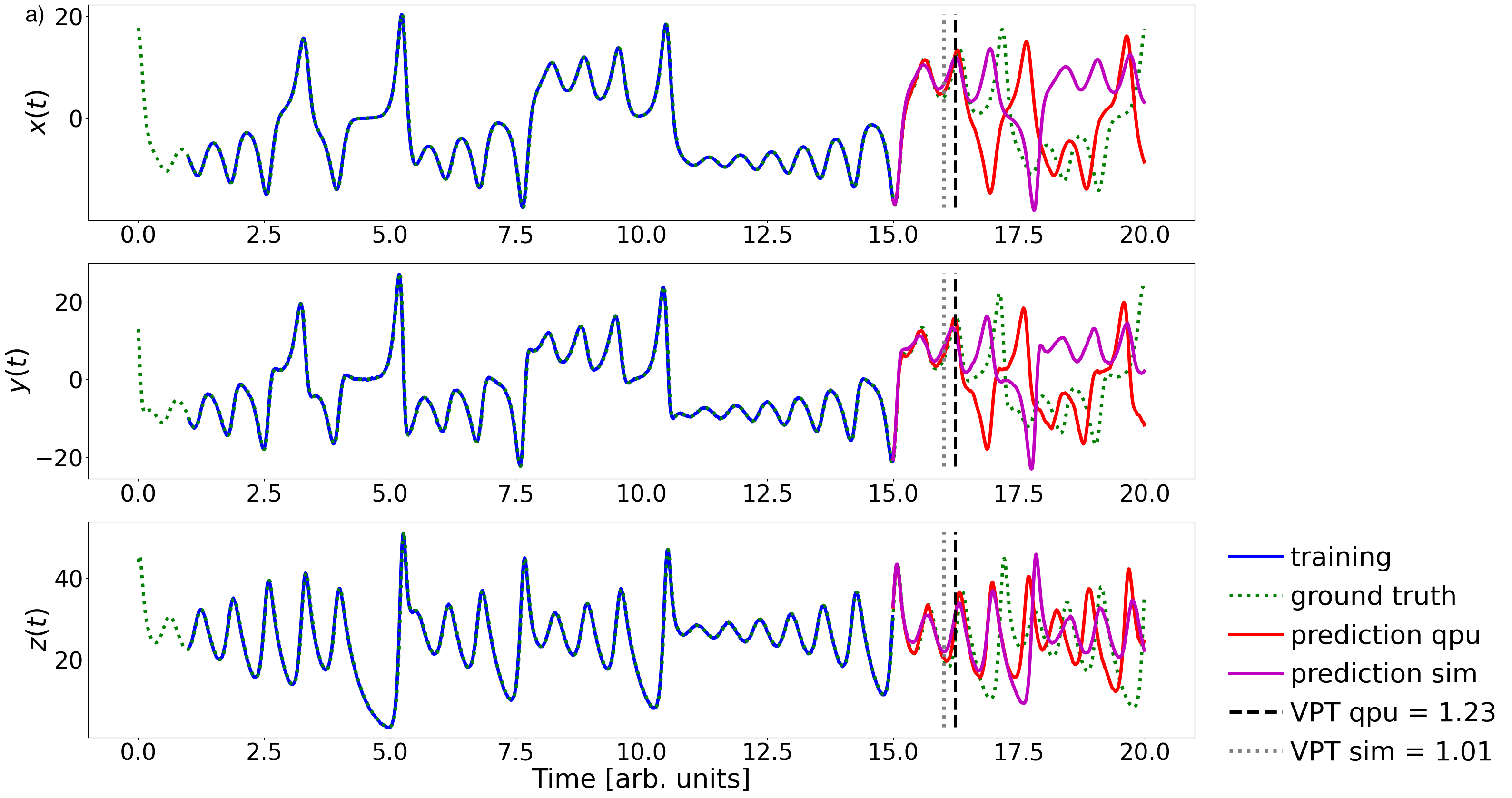}
    \includegraphics[width = 0.98\textwidth]{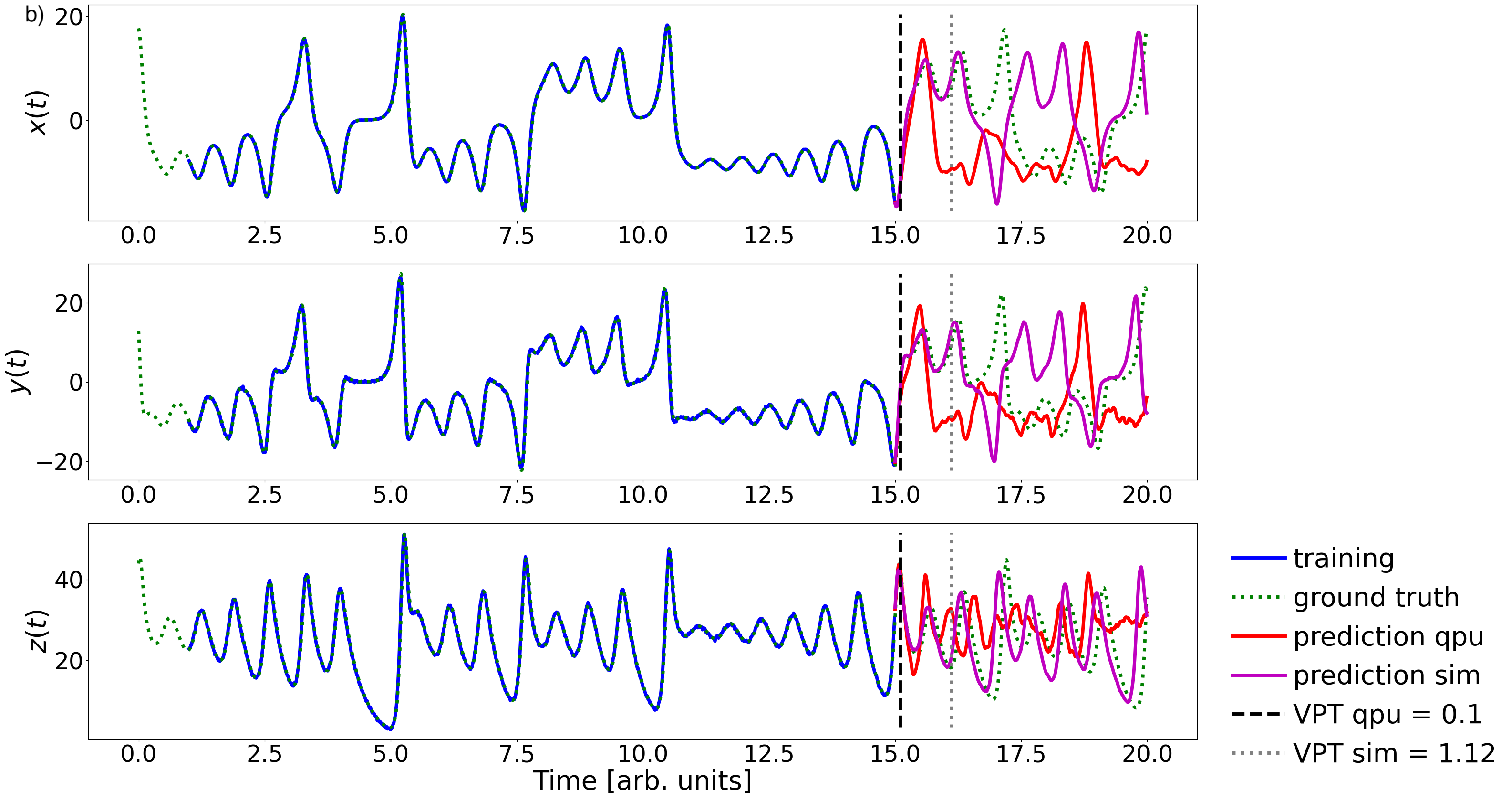}
    \caption{Comparison of 8-qubit QPU results from OQC Lucy chip (red lines) against noiseless classical simulations (purple lines) with 10,000 shots. a) Circuit without two-qubit gates,  with 6 single-qubit rotation gates acting on each qubit in the sequence of $R_X R_Y R_Z R_X R_Y R_Z$ with feature map $\phi(x) = \tanh(x)$. b) presents results for the same setups as discussed in Fig.~\ref{fig:best_result} (including two-qubit gates). }
    \label{fig:OQC_res}
\end{figure*}

We perform a proof-of-concept experiments on superconducting platform {\it Lucy} from Oxford Quantum Circuits (OQC), that comprises 8 qubit in ring topology. The presented results have two objectives: i) to identify feasibility of the HQRC on currently available platforms compered to the baseline of classical simulation, ii) to probe effects of real noise on the performance. Therefore, we restricted to two types of the HQRC implementation. One that has been discussed in \ref{sec:classical_L63}, i.e. 8 qubit case with two layers of type $L_1$ and 8 qubit case with only single-qubit gates (i.e. circuit is incapable of generating entanglement), both setups utilizing measurement vectors with single- and two-qubit expectation values. The former setup is exposed to higher level of hardware noise, as two-qubit gates display lower fidelities, while the latter case operates on reasonably high fidelities ($\ge$99.6\%). We present results in Fig.~\ref{fig:OQC_res}, which are obtained with 10,000 shots. For the low-level noise runs (only single-qubit gates) we observe a good qualitative and quantitative agreement with simulations resulting in marginally better VPT predictions, which are associated with shot-noise fluctuations and random disturbances of quantum dynamics due to hardware imperfections. Since, these imperfections in the latter case are deliberately suppressed by avoiding two-qubit gates and deep circuits, the HQRC algorithm can {\it learn} the native noise and adjust predictions. However, once the two-qubit gates are introduced to the system, noise is strong enough to disturb the observations, leading to short predictions and qualitative deviations shortly after the predictions have started (form ~16.0 time units).  These proof-of-concept results provide initial insights into real hardware implementation, and what needs to be the focal point if one intents to push the envelop of the QPU implementation. Therefore, circuit optimization and exploitation of native gates and connectivity, as well as deliberate selection of qubits based on hardware calibration (gate fidelities, read-out errors, $T_1$, $T_2$ and other metrics) is crucial step in recovering noiseless results. Additionally, employment of error-mitigation strategies can benefit the implementation, which was beyond the scope of this contribution. 

\section{Conclusions}

We have introduced the hybrid quantum reservoir computing (HQRC) model, an extension of classical reservoir computing that introduces additional complexity through the addition of measurements from a modular quantum circuit. These proof-of-concept results demonstrate that the HQRC algorithm is an interesting candidate as a model for short-term forecasting of chaotic time-series, being capable of reconstructing short-term predictions of both Lorenz63 and double-scroll chaotic systems, as well as providing correct long-term attractor behavior. This is despite the reservoir states in HQRC being lower in dimension than in the classical RC benchmarks (\textit{e.g.}in \cite{platt2022systematic} reservoir of size above 1000 yield VPT$>10$, while reservoirs of size $\sim 100$ can reach VPT$\sim 5$ in \cite{platt2022systematic} or $\sim 8$ for our classical RC approach). In particular, the classical RC state-of-the-art approaches rely on two-step optimization, where the additional optimization routine is to determine the most suitable set of hyperparamters. Understanding the proper strategy for hyperparameter tuning, however, requires further development, as the most crucial hyperparameters are related to the quantum Ansatz (\textit{e.g.} selection of number and type of layers can have different impact than fine-tuning regularization or leak rate values - we present extensive analysis based on sweeping the parameters in Appendix~\ref{app:layers}).. 

We observe that our method as implemented is sensitive to (even coherent) noise and finite number of samples. Additionally, the noise effects are even more amplified in case of proof-of-concepts results recovered from {\it Lucy} chip. We believe, that one can overcome these deficiencies with appropriately tailored classical transformations and error mitigation techniques. We leave this for future research. 

We finally note that the presented results are based on small system sizes (number of qubits $\neq10$) which can be easily simulated on a laptop. Therefore, the proposed algorithm, can also serve as a purely classical method. However, the simulation cost of full quantum circuits for the training and prediction phases is substantially more expensive than running state-of-art classical RC at this moment, so we do envision a hardware implementation. 

\section*{Acknowledgments}
We are grateful to the OQC team for their support and access to the {\it Lucy} chip. We appreciate stimulating discussions with Thiparat Chotibut, Supanut Thanasilp, Zo\"e Holmes and Osama Ahmed. 

\appendix

\section{Tested layers}\label{app:layers}
\begin{figure*}[!htbp]
    \centering
    \includegraphics[width = 0.98\textwidth]{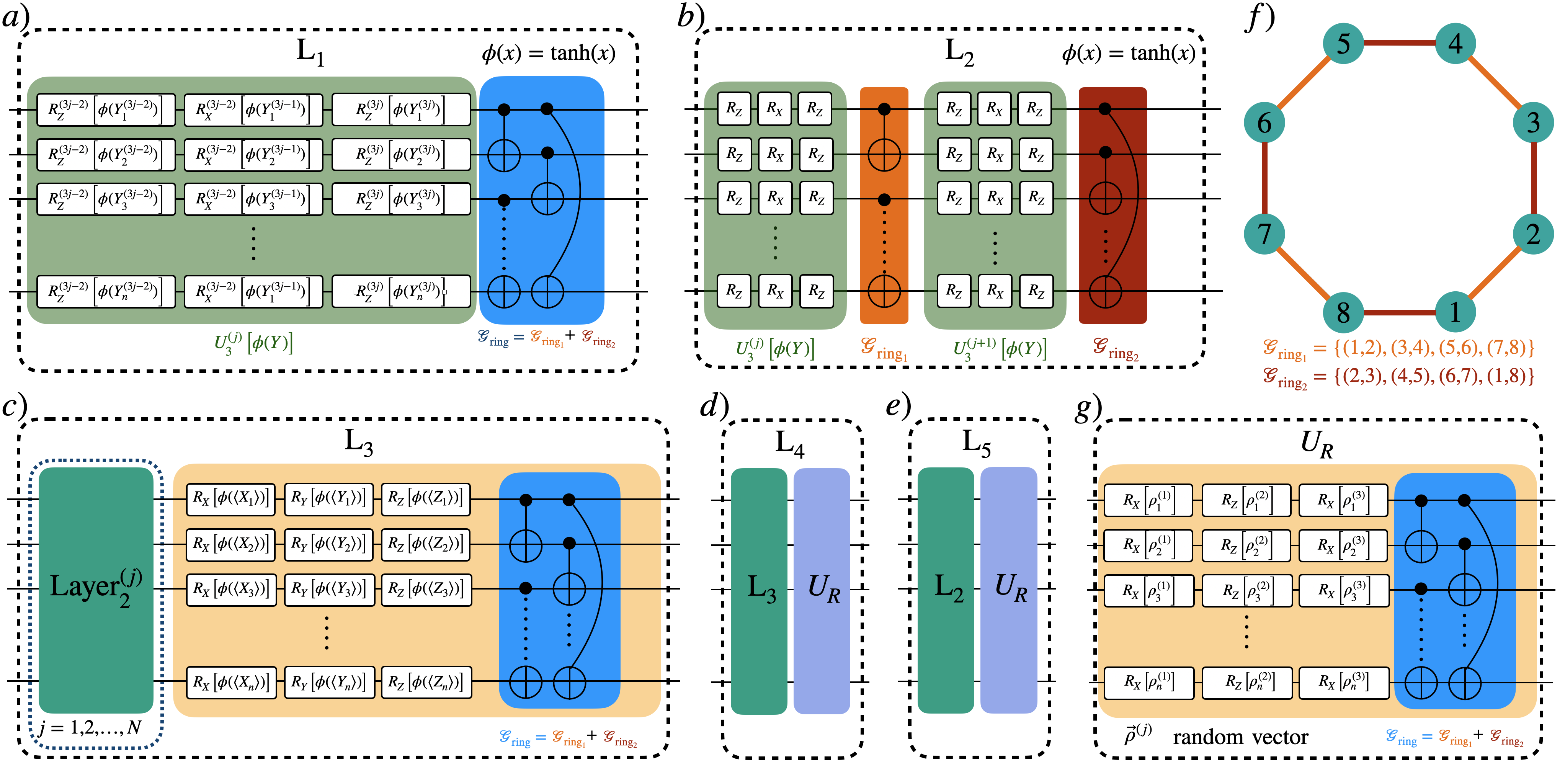}
    \caption{Main layer types that we explored for benchmarking Lorenz63 system. In a) the layer utilizes three single-qubit rotations parameterized by a feature map function $\phi$ acting on the transformed input $Y^{(j)} = W_{in}^{(j)} X_t$, where the components of the transformed vector are distributed from top to bottom in qubit rotations. Subsequently, we use a network of CX gates that are arranged in a ring graph, first acting according to $\mathcal{G}_{\mathrm{ring}_1}$, then as $\mathcal{G}_{\mathrm{ring}_2}$. An 8-qubit graph partitioning is depicted in f). The second type layer $L_2$ in b) has similar structure with even layers followed by a CX network on $\mathcal{G}_{\mathrm{ring}_1}$ and odd by $\mathcal{G}_{\mathrm{ring}_2}$ graph (here two layers are depicted). Panel c) displays a {\it measurement feedback} layer, which has $N$ layers of $L_2$ type followed by a measurement feedback block composed of single-qubit rotations $R_P(\langle P_i\rangle)$, where $P=X,Y,Z$ acting on $i$-th qubit. The expectation values are utilized from the previous time-step measurement $t-1$. The d) layer is a combination of $L_3$ layer and a random circuit, and panel e) has structure of layer $L_2$ followed by the random circuit. The random circuit has structure as in panel g), where three single-qubit rotations are parameterized by a random vector $\vec{\rho}$ drawn from a uniform distribution [0,2$\pi$], followed by a network of CX gates.  Note, that single-qubit expectation values of Pauli operators are fed with extra functional transformation $\phi$ into single-qubit rotations along the same axis as the operator.}
    \label{fig:layer_circuits}
\end{figure*}

On the high-level, the HQRC enables an arbitrary structure of the used Ansatz. Altough, it is beneficial to tailor it, such that it has sufficient expressibility to differentiate subtle differences in a problem to be solved. However, in most cases, this is {\it a priori} a complex task, and requires further systematic investigation on case by case basis. Here we present layers that have been tested, and allowed us to restrict the Ansatz to reasonably well-performing architectures. In particular, we have limited our search to repeated layers of general qubit rotations
\begin{equation}
    U_3(\alpha, \beta,\gamma) = R_Z(\gamma)R_X(\beta)R_Z(\alpha),
\end{equation}
followed by a network of CX gates with control and target qubits defined by a graph. Note, that $U_3$ exploits two $R_Z$ gates, that in the most hardware architecture are performed virtually without any additional cost. Furthermore, we focused on graphs that are fairly sparse or reflect connectivity of currently available devices  (see Fig.~\ref{fig:layer_circuits} f) panel). The main layers that we used are depicted in Fig.~\ref{fig:layer_circuits}. In addition to the gate allocation, these layers exploit feature maps in the form of a function $\phi(\cdot)$ transforming rotation angles (see Appendix~\ref{app:feature_maps} for further discussion). 

\begin{figure*}
    \centering
    \includegraphics[width = 0.31\textwidth]{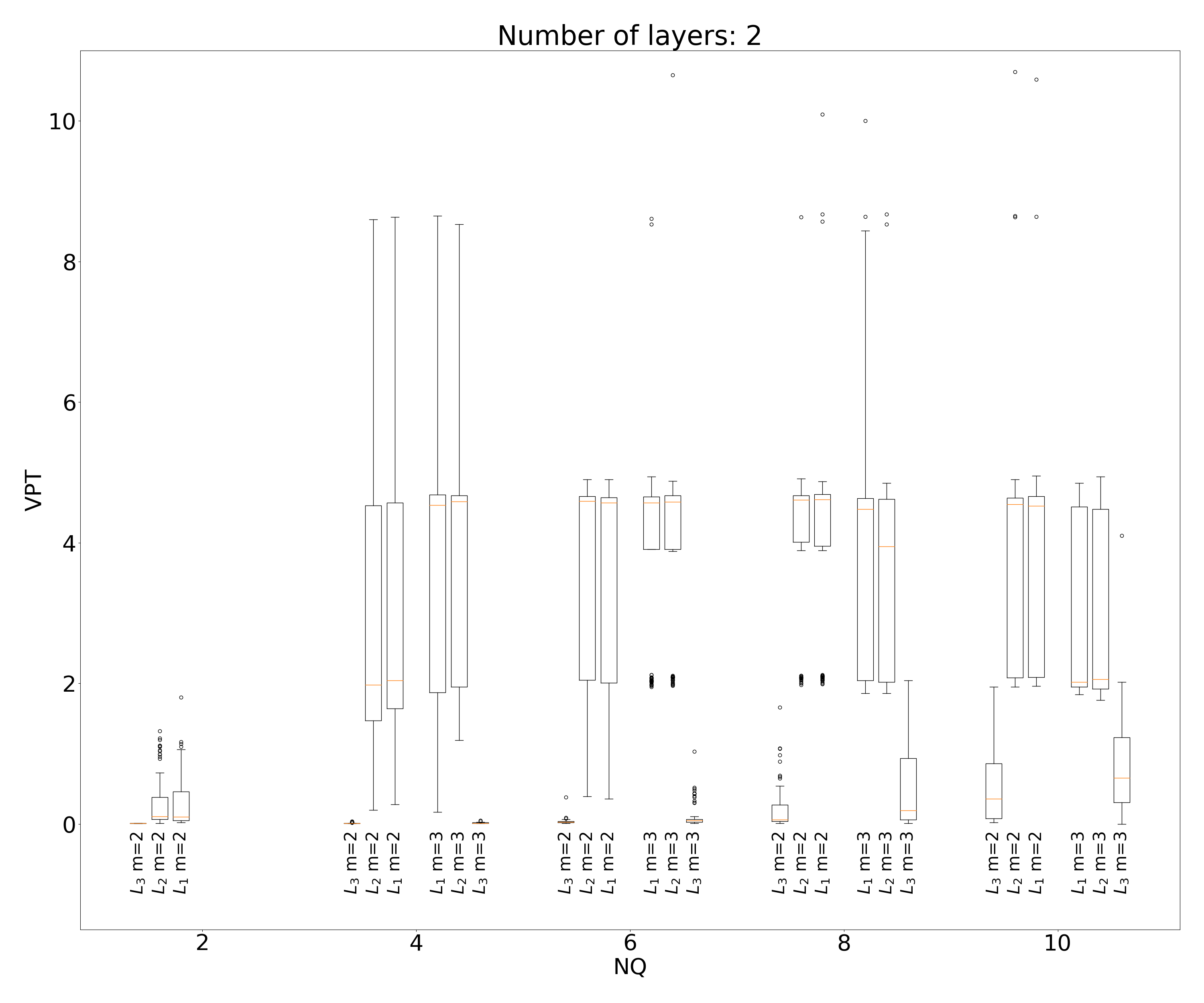}
    \includegraphics[width = 0.31\textwidth]{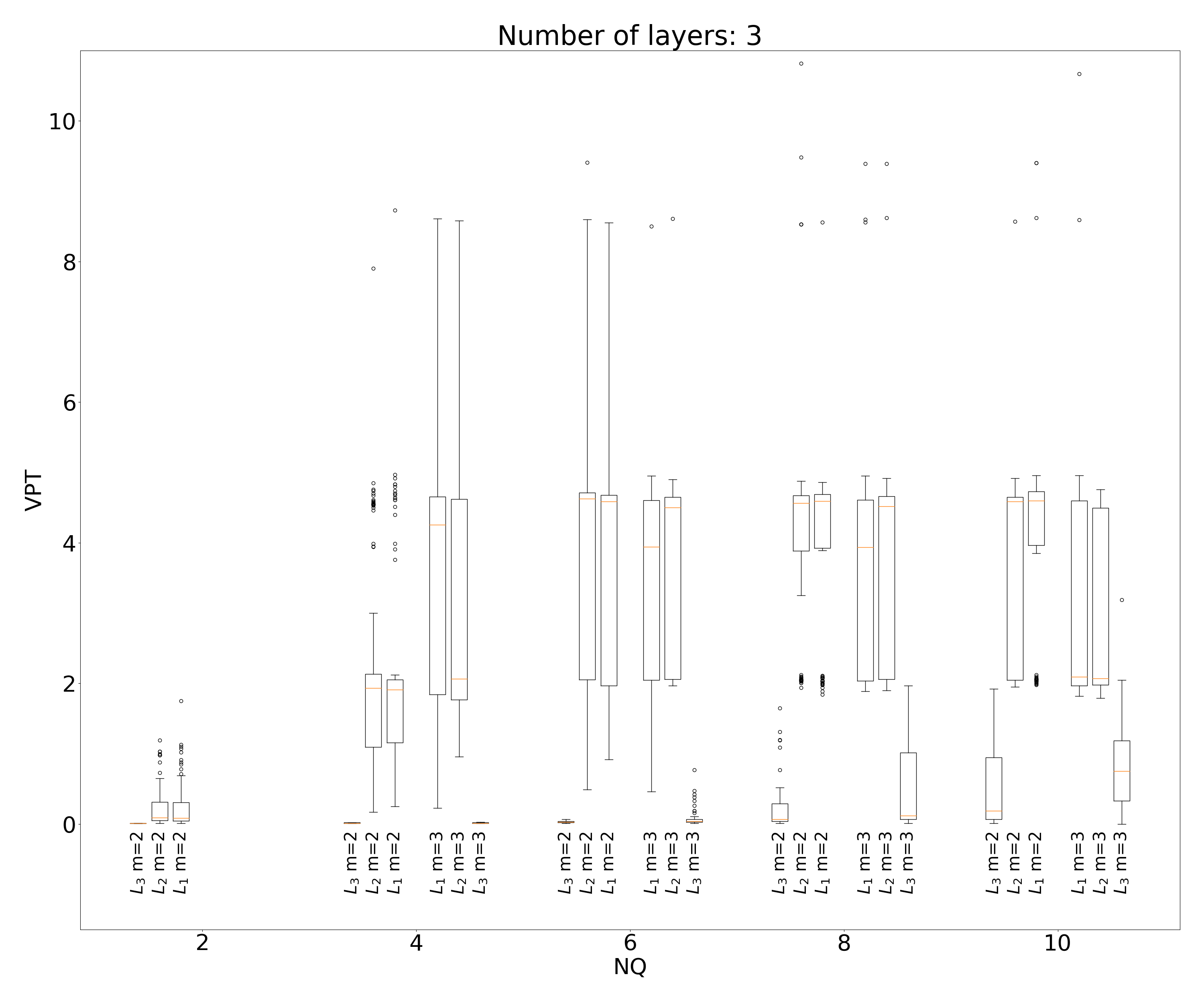}
    \includegraphics[width = 0.31\textwidth]{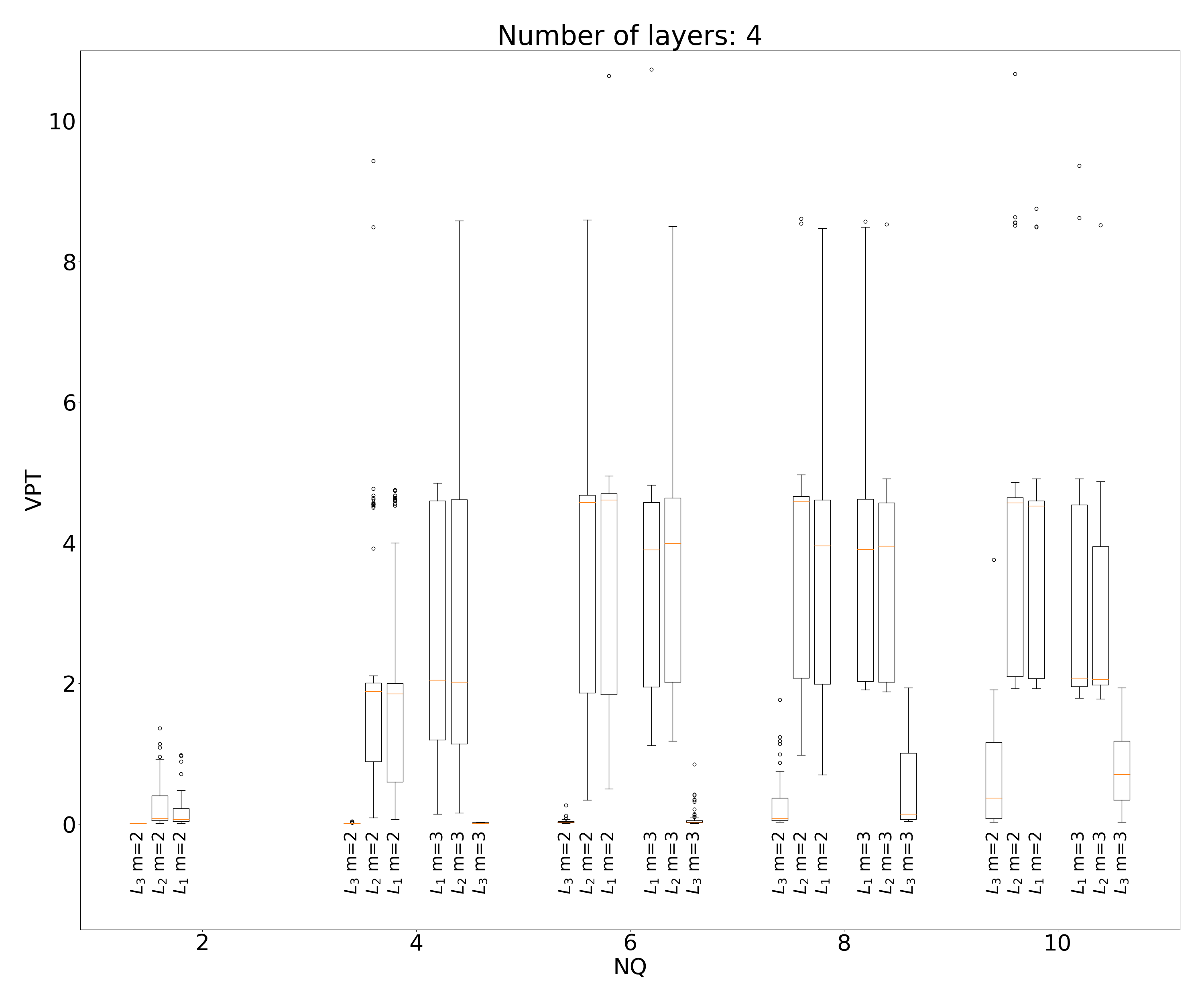}
    \includegraphics[width = 0.31\textwidth]{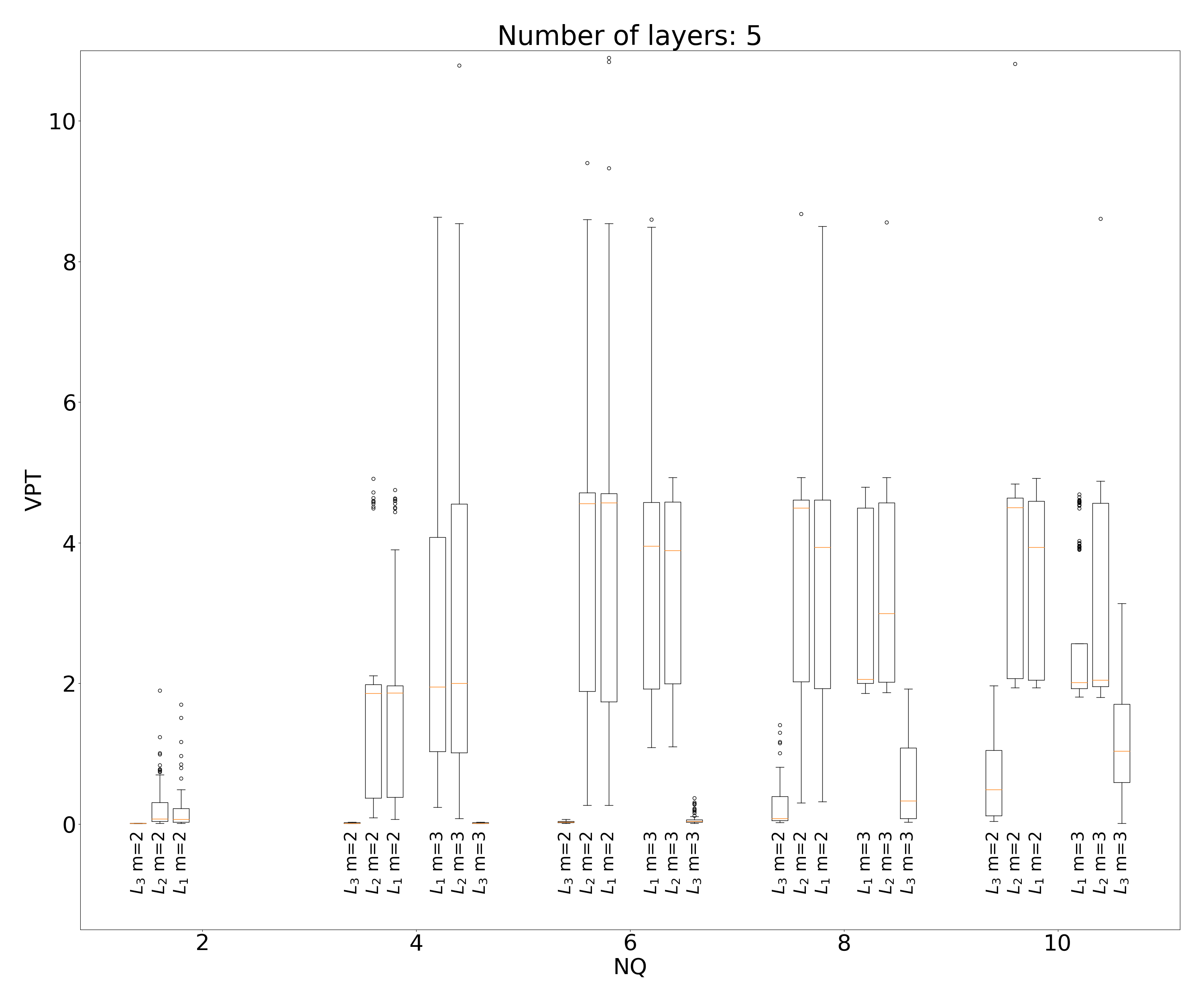}
    \includegraphics[width = 0.31\textwidth]{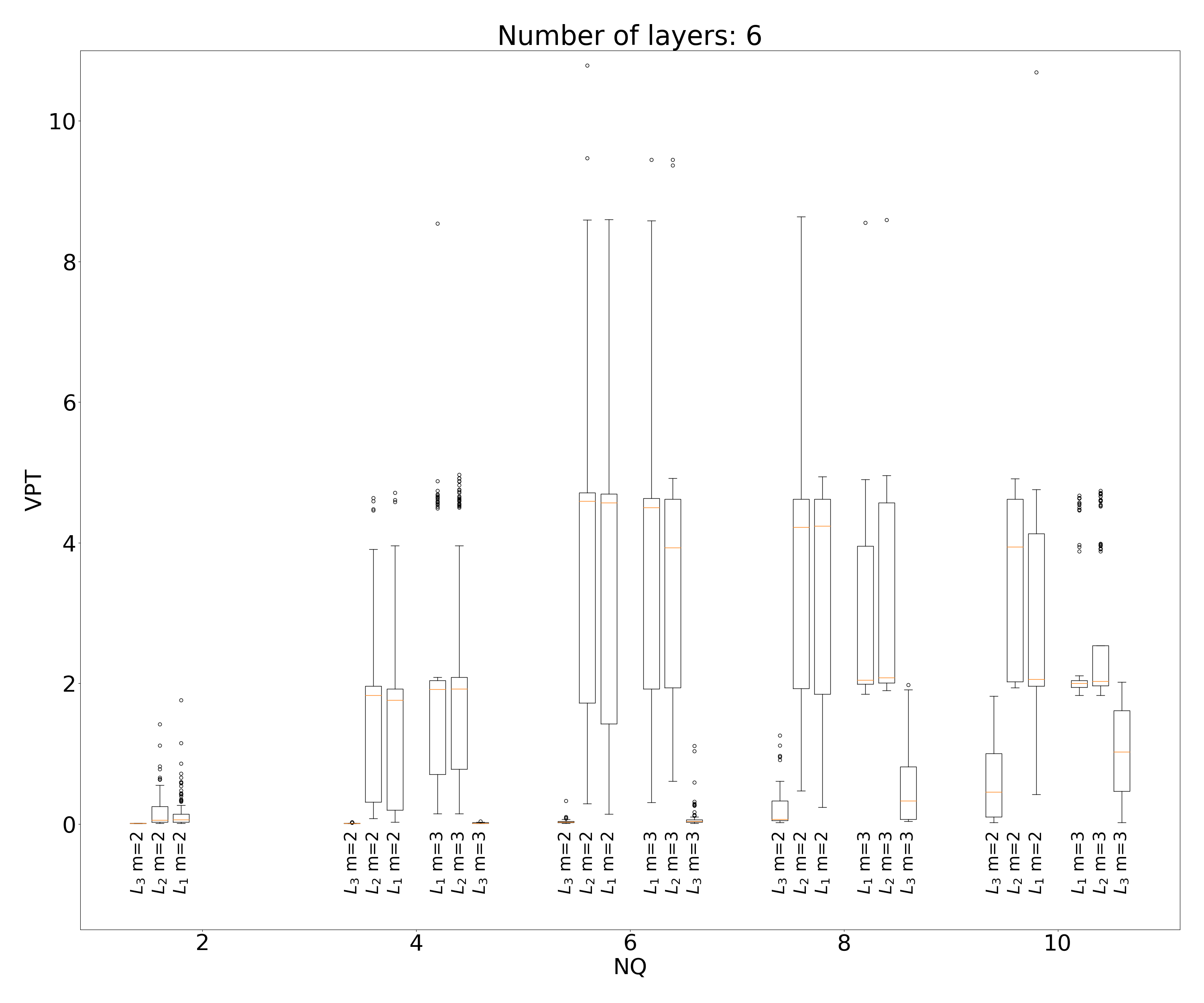}
    \includegraphics[width = 0.31\textwidth]{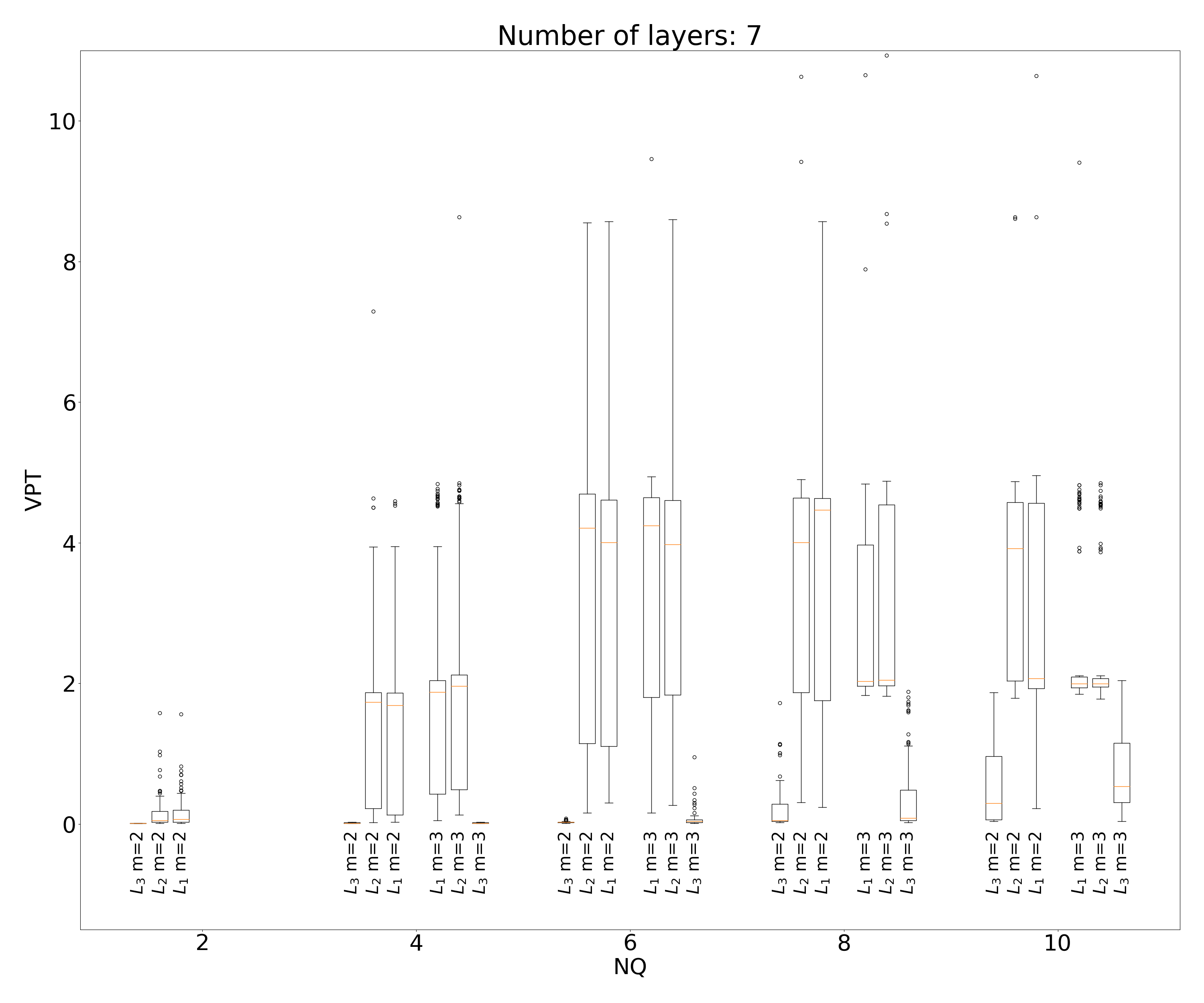}
    \caption{Box plot for different number of layers arranged along the increasing number of qubits used in the circuit. The data points are composed of 20 randomly selected transformation matrices, for 5 different feature map functions (see Appendix~\ref{app:feature_maps}). Each box contains data between first (Q1) and third (Q3) quantiles. The orange bar marks median value, while whiskers extend for Q1/Q3 $\pm$ IQR (interquantile range, i.e. IQR=Q3-Q1). The dots depict fliers that lay beyond whiskers. Each setup is labeled at the bottom of the figure with the type of the layer $L_1, L_2$ or $L_3$ and how many measurements have been used to compose the measurement vector ($m=3$ means that single, two- and three-body correlators are used, while $m=2$ contains up to second-order correlators). }
    \label{fig:box_fixed_layers}
\end{figure*}

\begin{figure*}
    \centering
    \includegraphics[width = 0.45\textwidth]{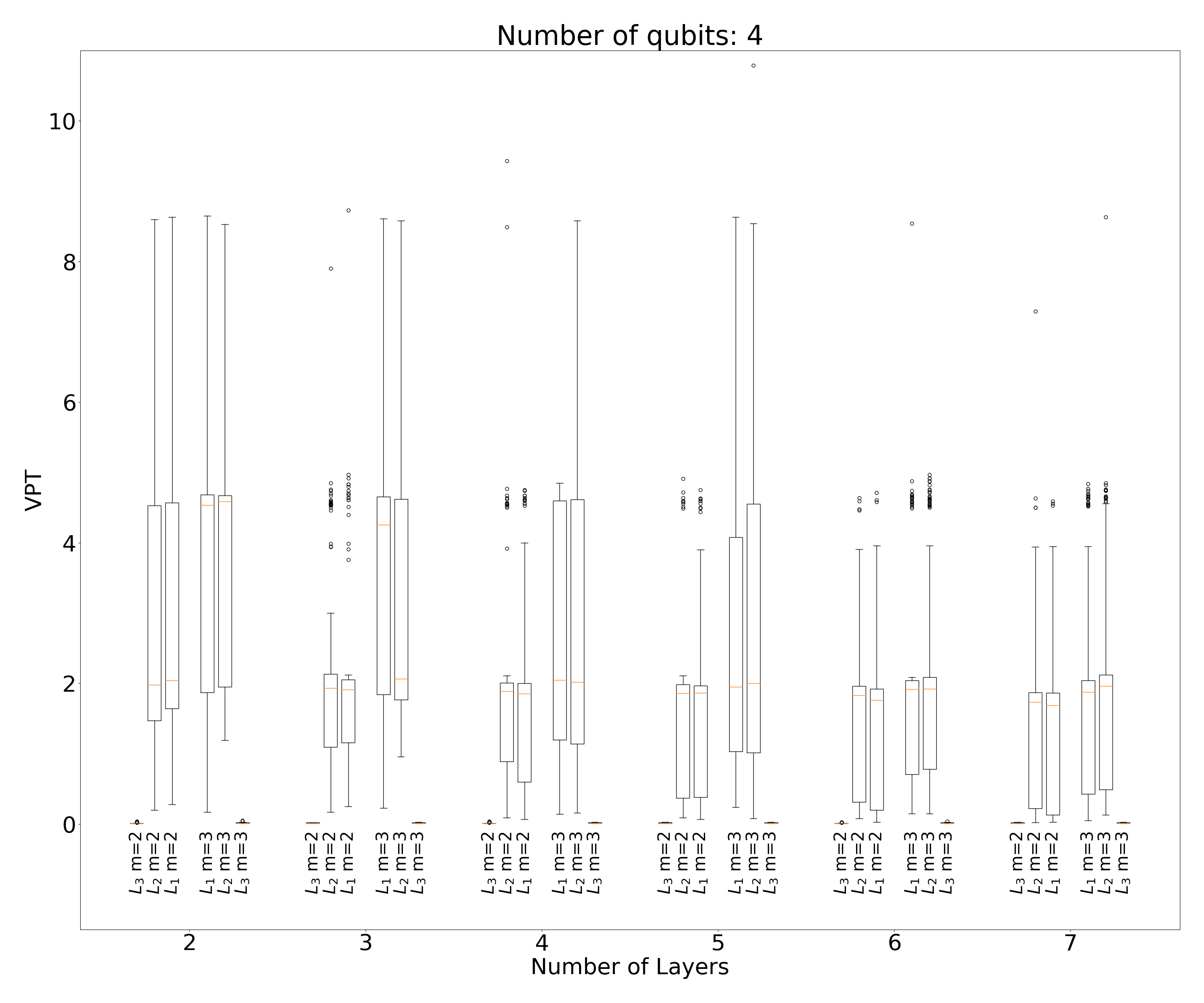}
    \includegraphics[width = 0.45\textwidth]{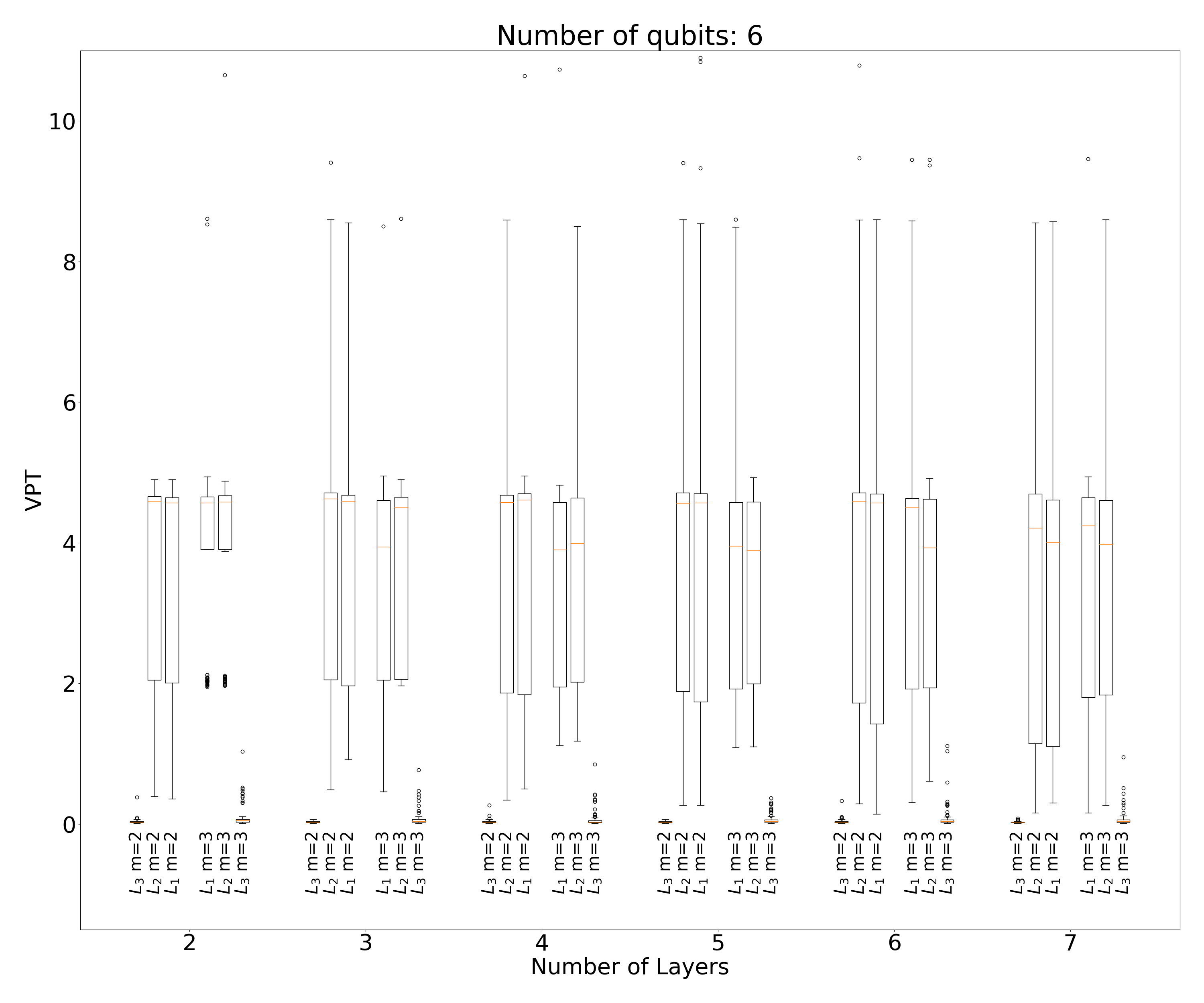}
    \includegraphics[width = 0.45\textwidth]{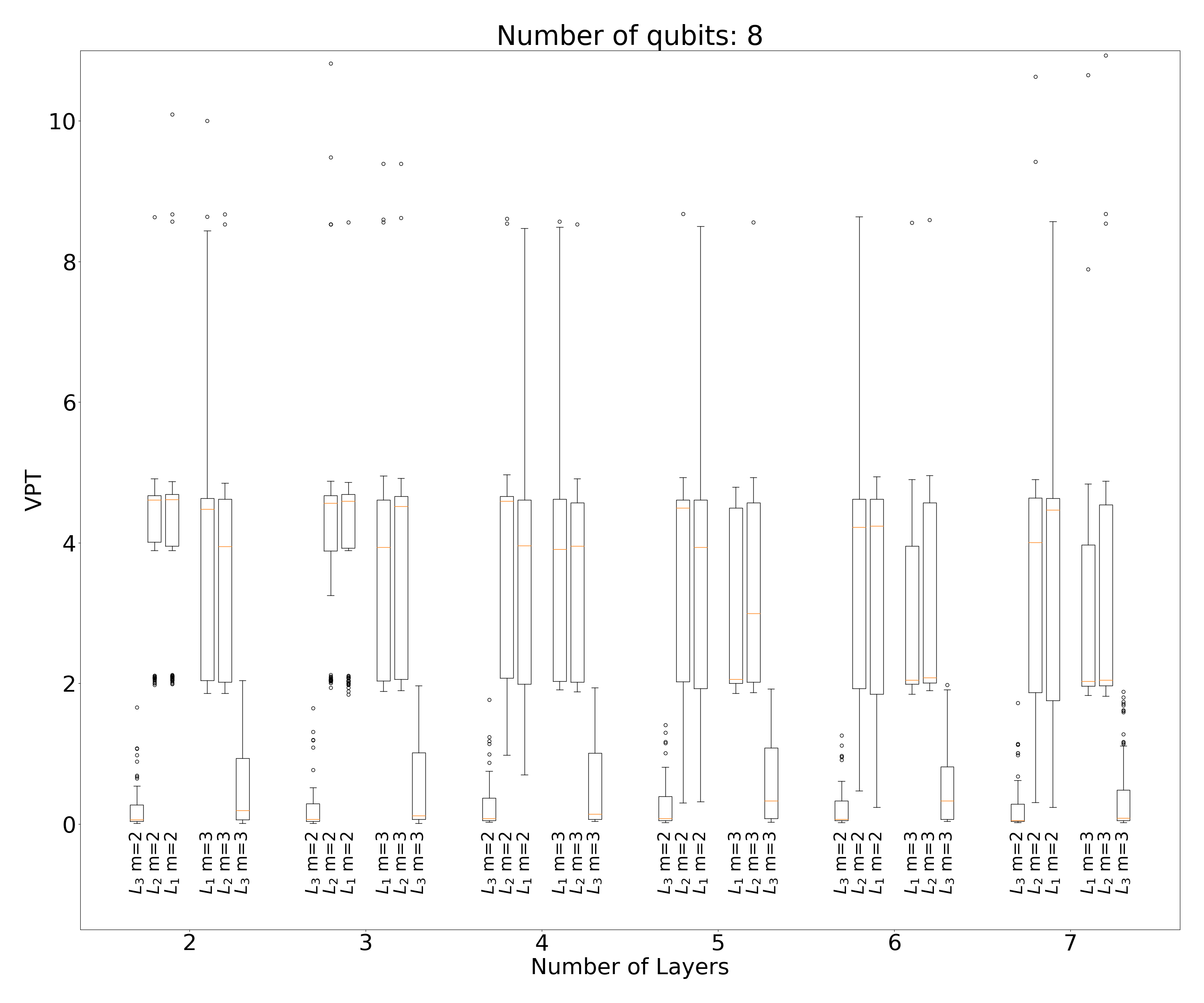}
    \includegraphics[width = 0.45\textwidth]{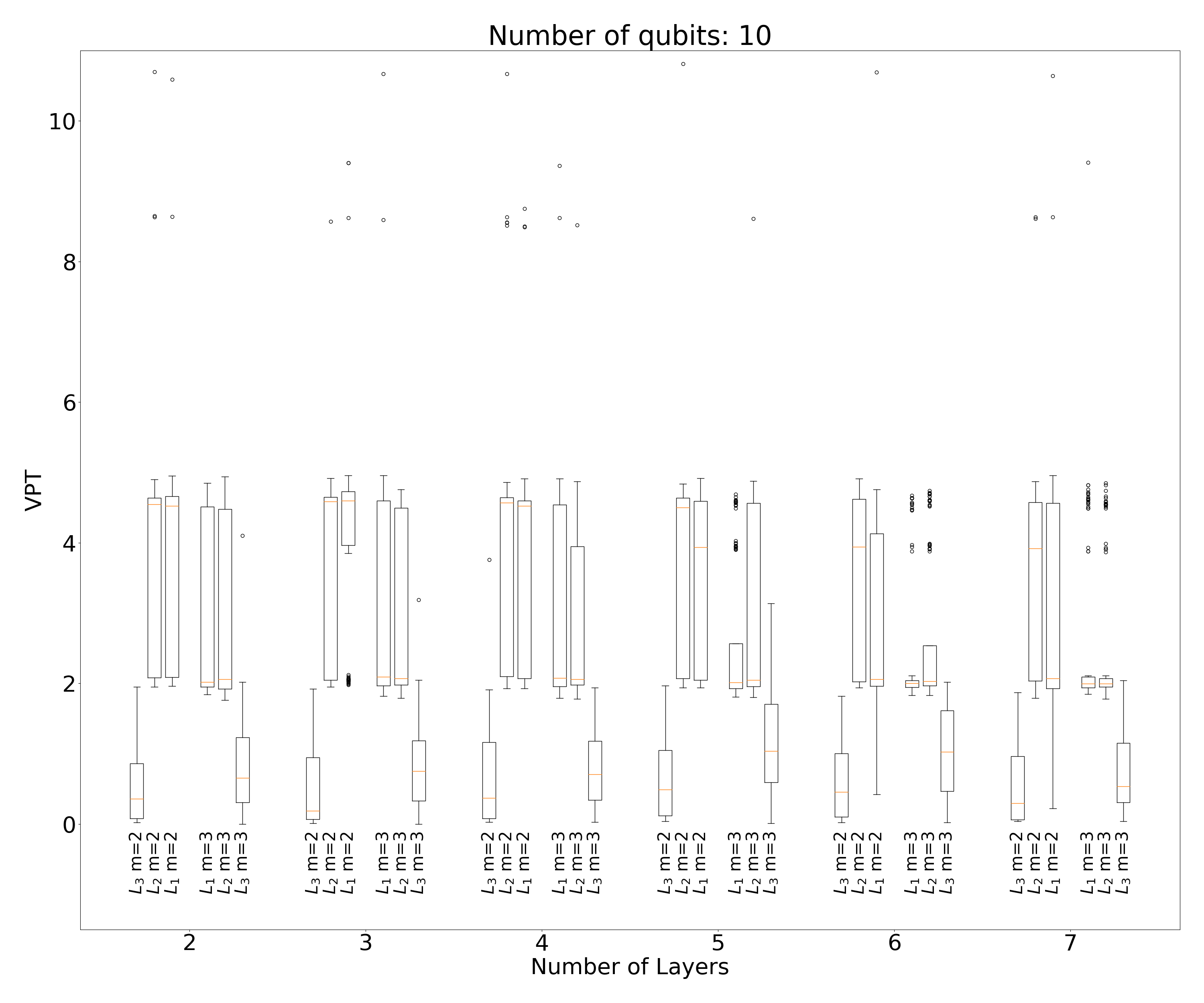}
    \caption{Box plot with similar box structure as in Fig.~\ref{fig:box_fixed_layers}, but now depicted for increasing number of layers with fixed number of qubits. }
    \label{fig:box_fixed_qubits}
\end{figure*}

In Fig.~\ref{fig:layers_stat} we compile statistical results for the Lorenz63 problem, where we test different layer types acting on different number of qubits, with different circuit depths (i.e. total number of layers) and utilizing different feature map functions. The presented results are for noiseless simulations with exact expectation value. The circuits for each layer type are depicted in the collection of figures that display VPT data as a function of number of qubits with fixed number of layers Fig.~\ref{fig:box_fixed_layers}, as a function of number of layers with fixed number of qubits Fig.~\ref{fig:box_fixed_qubits}, both sets of plots combine 20 random seed initialization of the network for each different feature map encoding (see Appendix~\ref{app:feature_maps}) and measurement vectors composed with up to second- and third-order correlators (total of 810 setups). This statistical sweep through different types of layers and circuit properties (depth, i.e. number of layers, and qubits) has been done for a fixed 1,500 training time steps with first 100 time steps discarded in the warm-up period (referred to as pruning length in reservoir computing nomenclature). We additionally fixed leak rate at $\alpha =0.7$ and regularization to $\beta = 10^{-8}$. 

The results display large variability within the setups. One can identify that increasing number of qubits, on average, result in better performance, however this trend is nondeterministic and one can experience larger values in VPT for fewer qubits. On the other hand, for Lorenz63, we fail to observe clear improvements with increasing number of layers, suggesting that sufficient information scrambling and extraction of relevant properties for the quantum state can be achieved even for shallow circuits. This observation, in order to be conclusive, would need to be tested on higher dimensional systems of larger complexity, where appropriate distribution, and separation of relevant dynamical signatures is required. Additionally, both types of figures (Fig.~\ref{fig:box_fixed_layers} and \ref{fig:box_fixed_qubits}) cluster most of the results in the range of VPT between 4 and 6. 

\begin{figure}
    \centering
    \includegraphics[width= 0.45\textwidth]{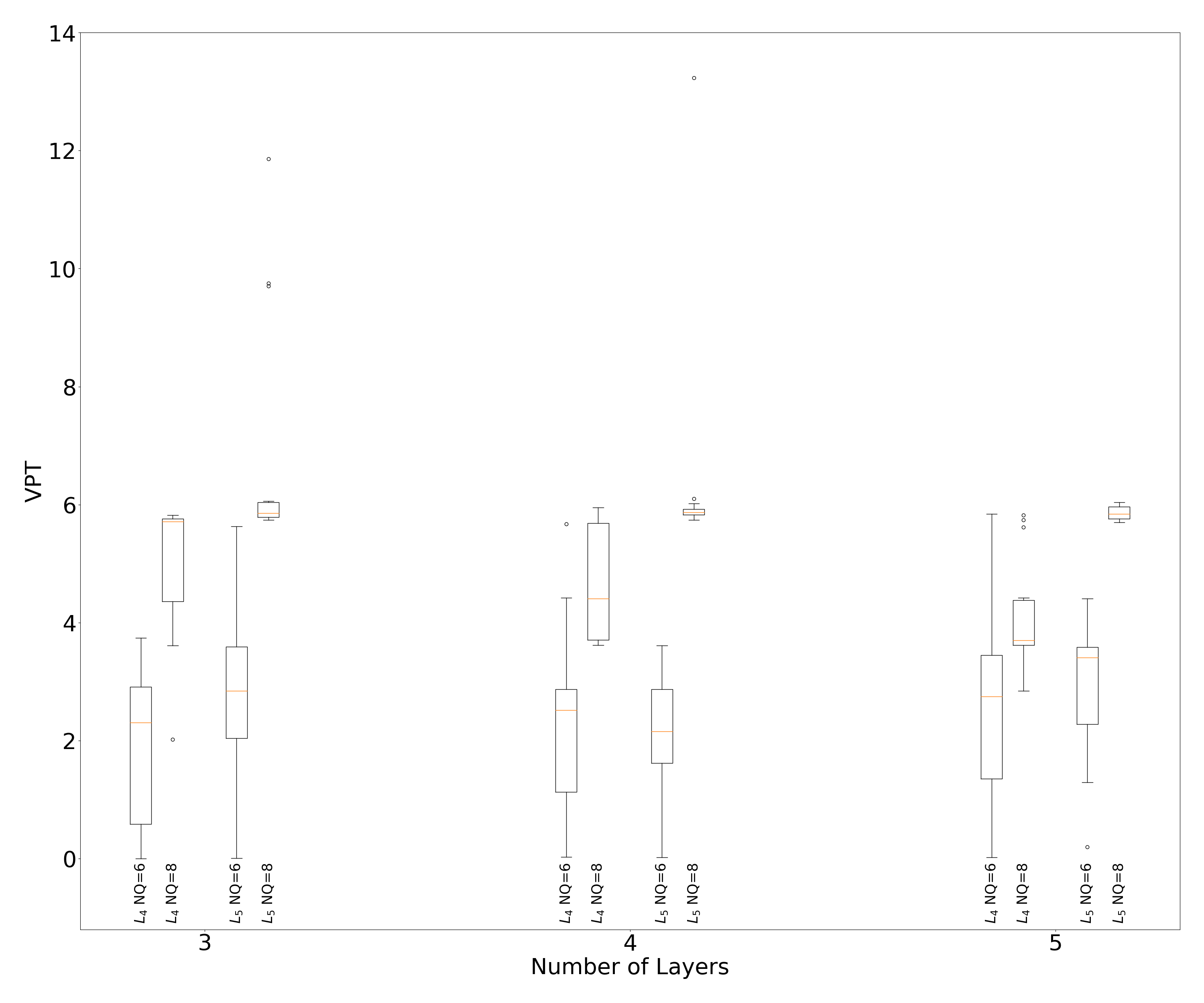}
    \caption{Box plot with the same box structure as in Fig.~\ref{fig:box_fixed_layers} depicting VPT values for statistical data for $L_4$ and $L_5$ layers  with 2,000 training steps, 100 prune length, $\alpha =0.7$, $\beta=10^{-8}$, $\phi(x)=\pi\sigma(x)$ and up to second-order correlators for creating the measurement vector.}\label{fig:L4L5}
\end{figure}

Based on the above results, we additionally tested $L_4$ and $L_5$ layers from Fig.~\ref{fig:layer_circuits} for networks with 6 and 8 qubits, and number of layers 3, 4 and 5. We fixed number of training steps to 2,000, prune length to 100, leak rate $\alpha = 0.7$ and regularization $\beta = 10^{-8}$ and $\phi(x) = \pi\sigma(x)$ feature map function and limiting to up to second-order correlators\footnote{Different hyperparameters settings than  in $L_1$-$L_3$ layers have been chosen based on results from Fig.~\ref{fig:box_fixed_layers}, Fig.~\ref{fig:box_fixed_qubits} and ones from Appendix~\ref{app:hyperparams}.}.  The results are depicted in Fig.~\ref{fig:L4L5}. This setup display manifestly better VPT values for 8 qubits, and unclear trend with respect to the number of layers, as in the case of $L_1$-$L_3$ layers.  Additionally, introduction of the random circuit layer after the measurement feedback one, helps to generate significantly better results in comparison to $L_3$ (i.e. without the random part), though still inferior to $L_5$. 
Based on the Lorenz63 data, and tested layers we identified $L_1$ layer as arguably the best performing setup, however, the best VPT value of 13.23 is obtained for $L_5$ type with 4 layers acting on 8 qubits using $\phi(x) = \pi \sigma(x)$ feature map, where $\sigma(x)$ is the sigmoid function (see Appendix~\ref{app:feature_maps}). Layers containing measurement feedback, demonstrated low VPT values, suggesting that direct incorporation of measurement outcomes fails to benefit the performance, regardless of the number of qubits, layers or type of feature map. On the other hand, inclusion of the random circuit at the end, can improve performance (at least if focusing on the outliers) at the expanse of deeper circuits.

\subsection{Feature maps}\label{app:feature_maps}
As appropriate selection of feature maps can impact the performance, here we restrict to a handful of choices inspired by earlier works \cite{Mitarai_2018,Kyriienko_2021,Schuld_2019}, we use $\phi$ in layers in Fig.~\ref{fig:layer_circuits}. We tested feature maps of the form:
\begin{enumerate}
    \item $\phi(x) = \tanh(x)$,
    \item $\phi(x) = \pi \tanh(x)$,
    \item $\phi(x) = \pi \sigma(x)$, where $\sigma(x) = 1/[1+\exp(-x)]$ is the sigmoid function,
    \item $\phi(x) = x$,
    \item $\phi(x) = \pi x$.
\end{enumerate}
Since we employ normalization of the training data, such that it belongs to $[-1,+1]$, multiplication by $\pi$ factor expands that image of the feature maps to $[-\pi,+\pi]$, allowing us to utilize more expressivity from the single-qubit rotations. 

\section{OQC Lucy results}\label{app:OQC_res}
\begin{figure*}
    \includegraphics[width = 0.96\textwidth]{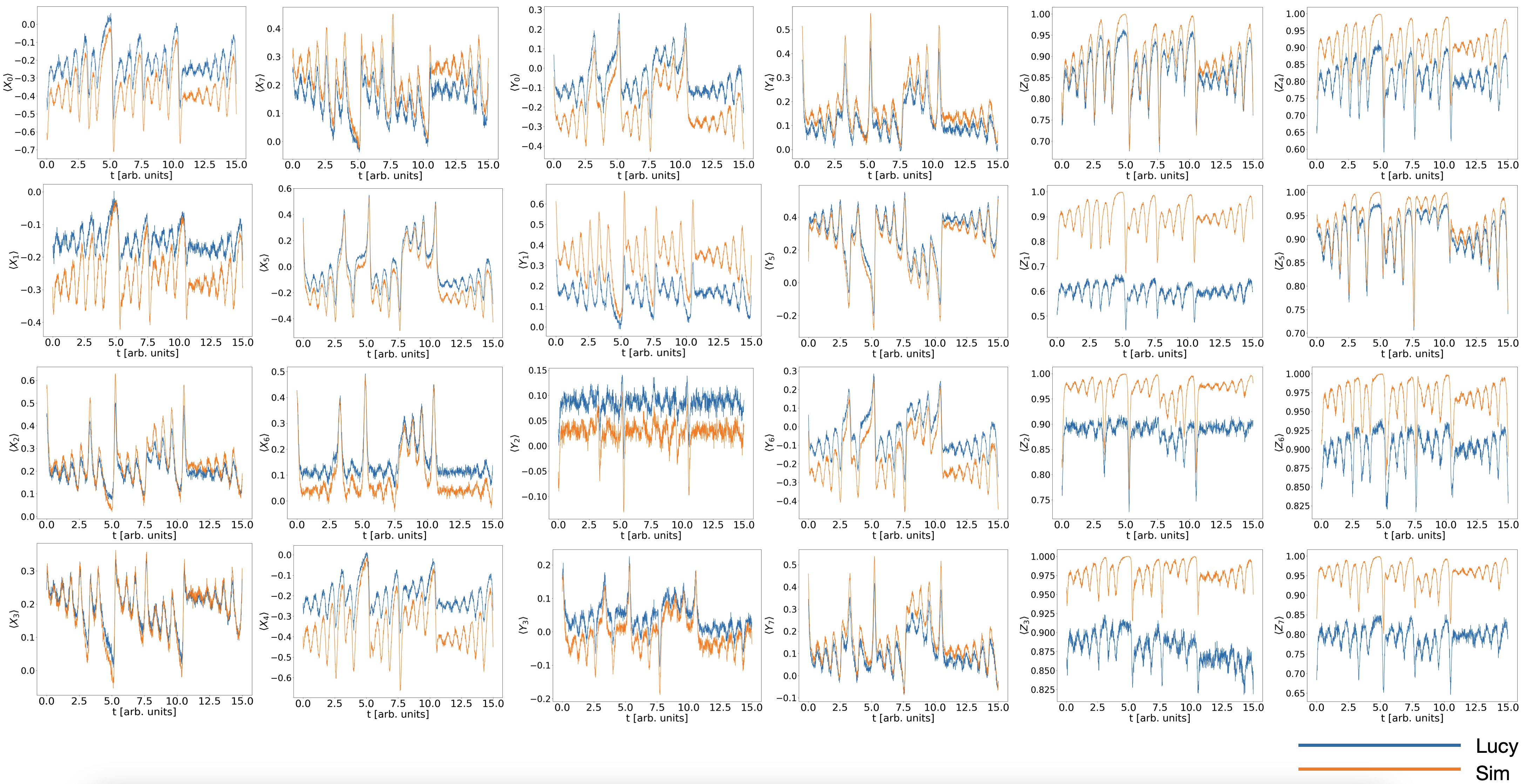}
    \caption{Comparison between simulation (orange) and QPU results from OQC Lucy chip (blue), both setups operating with the same hyperparameters and number of shots: 10,000. The steps is composed of only single-qubit rotations without two-qubits gates - same as one discussed in \ref{fig:1Q_meas}. The results display all single-qubit expectation values for three Pauli observables $X,Y$ and $Z$.}\label{fig:1Q_meas}
\end{figure*}

\begin{figure*}
    \includegraphics[width = 0.96\textwidth]{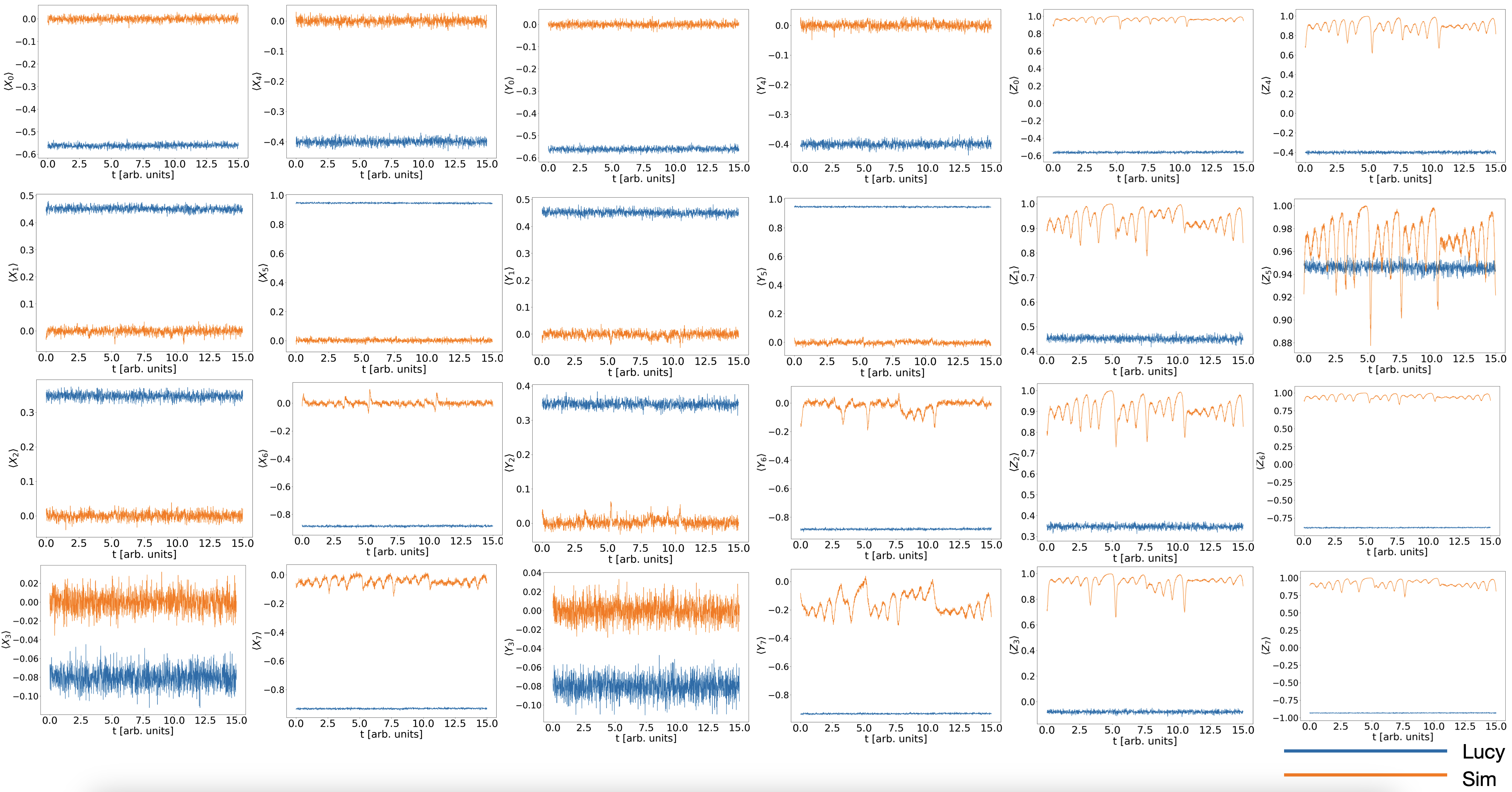}
    \caption{Comparison between simulation (orange) and QPU results from OQC Lucy chip (blue), both setups operating with the same hyperparameters and number of shots: 10,000. The setup is the same as in Fig.~\ref{fig:best_result} and is composed of 2 layers of CNOT interleaved by single-qubit rotations. The results display all single-qubit expectation values for three Pauli observables $X,Y$ and $Z$.}\label{fig:CX_meas}
\end{figure*}

Here we display supplemental information to the runs performed on OQC Lucy chip. In particular, we focus on measurement outcomes during training period period to understand how noise affects the performance. In~Fig.~\ref{fig:1Q_meas} and Fig.~\ref{fig:CX_meas} we compare noiseless simulations with QPU results (both with 10,000 shots) on the level of single-qubit expectation value (for circuits without entanglement generation, higher order correlators are simple products of single-qubit expectation values), as these provide sufficient information to understand observed deviations. 

The case without two-qubit gates in the circuit displays good qualitative agreement for all single-qubit expectation values, with some experiencing a clear upwards or downwards bias. Crucially, the general pattern is preserved, and the presence of the bias is not detrimental to the final observations, hence comparable VPT values (see~Fig.~\ref{fig:OQC_res} a)). In the case of circuits with two-qubit gates, many single-qubit expectation values carry close to zero distinguishable signal about the underlying dynamical system, and their profiles resemble white noise. This is true for both QPU and simulation results. The difference, however, is that for observables with visible fine structure in simulations, the QPU results still display white noise behavior. Additionally, the bias in these measurement outcomes is more pronounced than in the case without entangling gates. These two factors contribute to substantially lower VPT values, and faster qualitative divergence from simulation expectations. These observations motivate us to employ custom error mitigation strategies, and potentially explore adaptive approaches that could rely on measurement outcomes that strengthen predictions. However, we leave this for future research.

\section{Hyperparameters}\label{app:hyperparams}
In Table~\ref{tab:hp_global} we present all hyperparamters that are tunable in the algorithm, with some values that we fix in our simulations.

\begin{table*}[!htbp]
    \centering
    \begin{tabular}{m{4cm}|m{8cm}| m{4cm}}
        HP & description  & value  \\\hline\hline 
        $|\psi_{in}\rangle$ & initial state & $|0\rangle$\\  \hline 
        leak rate  & $\alpha\in[0,1]$,  controls contribution from the past reservoir state & $\alpha=0.7$\\\hline 
       seed  & fixes random matrices for reproduction purposes &  varying \\\hline 
        number of shots & how many times each circuit is measured to approximate expectation values & 1,000; 5,000; 10,000, 25,000, 50,000 or $\infty$ \\\hline 
        $f_R$ & a transformation function for the reservoir states $r_t$ in order to prepare $R_t$ training vector  & $\tanh$  \\\hline 
        $f_x$ & a transformation function for input data $X_t$ in order to prepare $R_t$ training vector  & $\tanh$ \\\hline 
         $f_X$ &a transformation function that acts on a transformed input data in order to create the reservoir state & identity \\\hline 
         $f_r$ & a transformation function that acts on the transformed previous reservoir state in order to create the current reservoir state & identity  \\\hline 
        $f_M$ & a transformation function that acts on a transformed measurement data in order to create the reservoir state & identity     \\\hline 
        $g$ & a transformation function that acts on contributions from the previous reservoir state, measurement outcomes, and input state in creation of the new reservoir state& identity\\\hline
        training length & how many time steps are used for training required for ridge regression  & 1,500 - 2,000   \\\hline 
        test length & how many time steps are predicted & 1200 \\\hline 
        prune size (warm up) & how many initial time steps are discarded from the training procedure & 100\\\hline 
        regularization $\beta$ & stabilizes matrix inversion by addition of $\beta I$ in ridge regression & $10^{-6}, 10^{-7},10^{-8}$\\\hline
        input type & what is the source of input for parametrized gates & data, measurement feedback, random\\\hline 
        gate & gate type in the layer  & $R_Y,R_Z,R_X, CX$ \\\hline 
        graph & graph of qubits on which the gates act  & $\mathcal{G}_{\mathrm{ring}}$   \\\hline 
        $\phi$ & feature map function that transforms input parameters & $\tanh$, $\pi\tanh$, $\pi\sigma$, $id$, $\pi\circ id$\\\hline
        Pauli basis & which measurement basis is selected & $X,Y$ and $Z$ \\\hline
        graph & which qubits constitute the measurement correlators  & all-to-all connectivity\\
        \end{tabular}
    \caption{A list of hyperparameters that specify the algorithm's setup. Here we focus on parameters related to the classical part of the algorithm. Value column refers to what are common choices in our experiments.}
    \label{tab:hp_global}
\end{table*}

It is crucial to understand what role is played by different hyperparamters. Here we collect a number of results testing VPT performance of the algorithm. We select the HQRC with $\phi(x) = \pi\sigma(x)$ feature map implemented on 6 qubit circuit with four $L_2$ layers, and we use up to third-order correlators. For this section's analysis we fixed the random seed. 

\begin{figure}[!htbp]
    \centering
    \includegraphics[width = 0.45\textwidth]{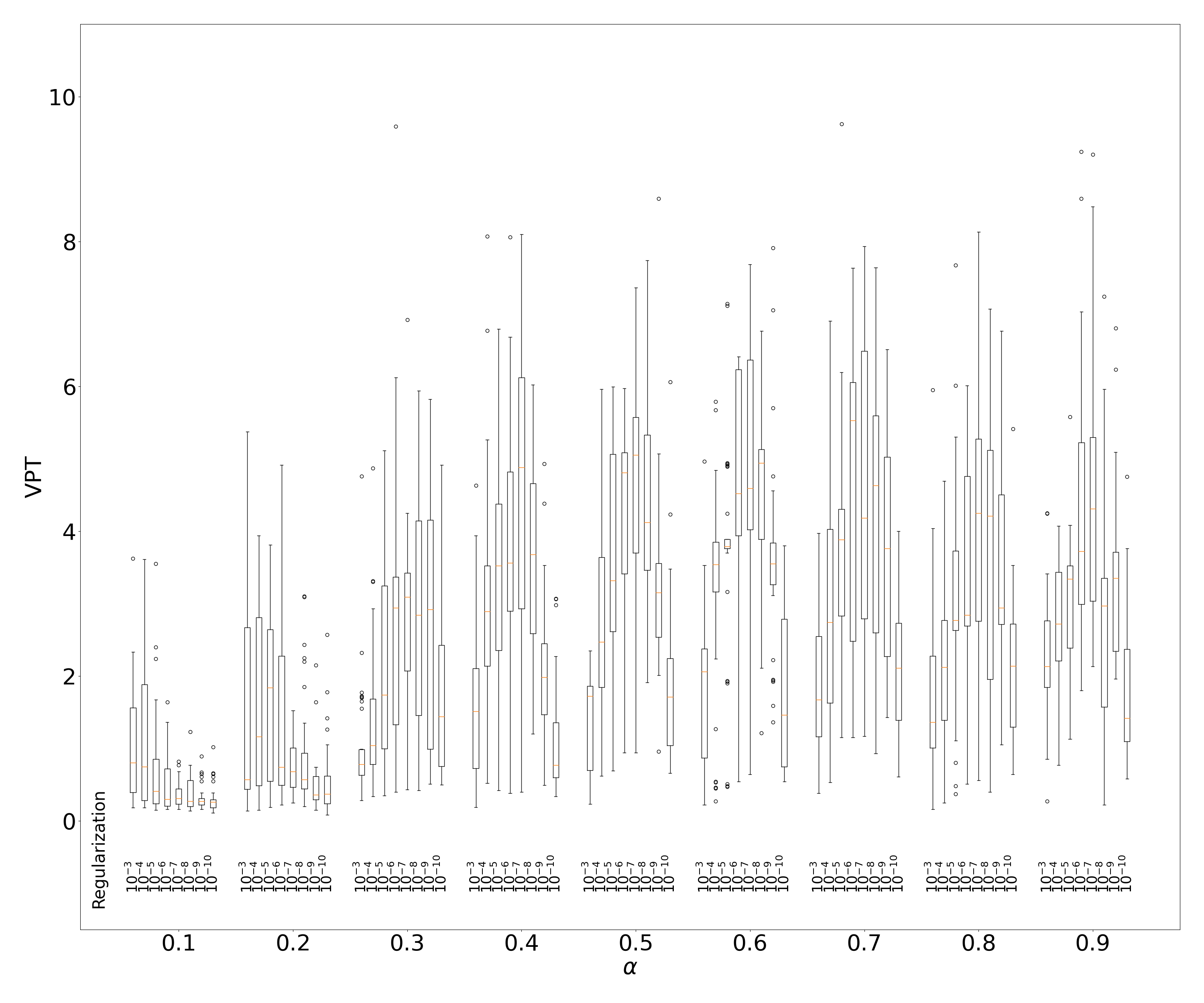}
    \caption{Box plot with the same box structure as in Fig.~\ref{fig:box_fixed_layers} depicting VPT values for statistical data for different values of leak rate $\alpha$ and regularization $\beta$. }
    \label{fig:reg_leak}
\end{figure}

In Fig.~\ref{fig:reg_leak} we test the impact of regularization strength and leak rate. Regularization can help stabilize results and prevent overfitting during the training phase, while leak rate controls the amount of past information that is retained in the network (the larger $\alpha$ is the smaller impact previous iterations have).

We see that optimal values of leak rate are between 0.4-0.8 with regularization $10^{-6}-10^{-8}$. Smaller and larger regularizations tend to fail stabilize the results, and the optimal region, additionally display outliers shifted towards higher values of VPT. 

\begin{figure}[!htbp]
    \centering
    \includegraphics[width = 0.45\textwidth]{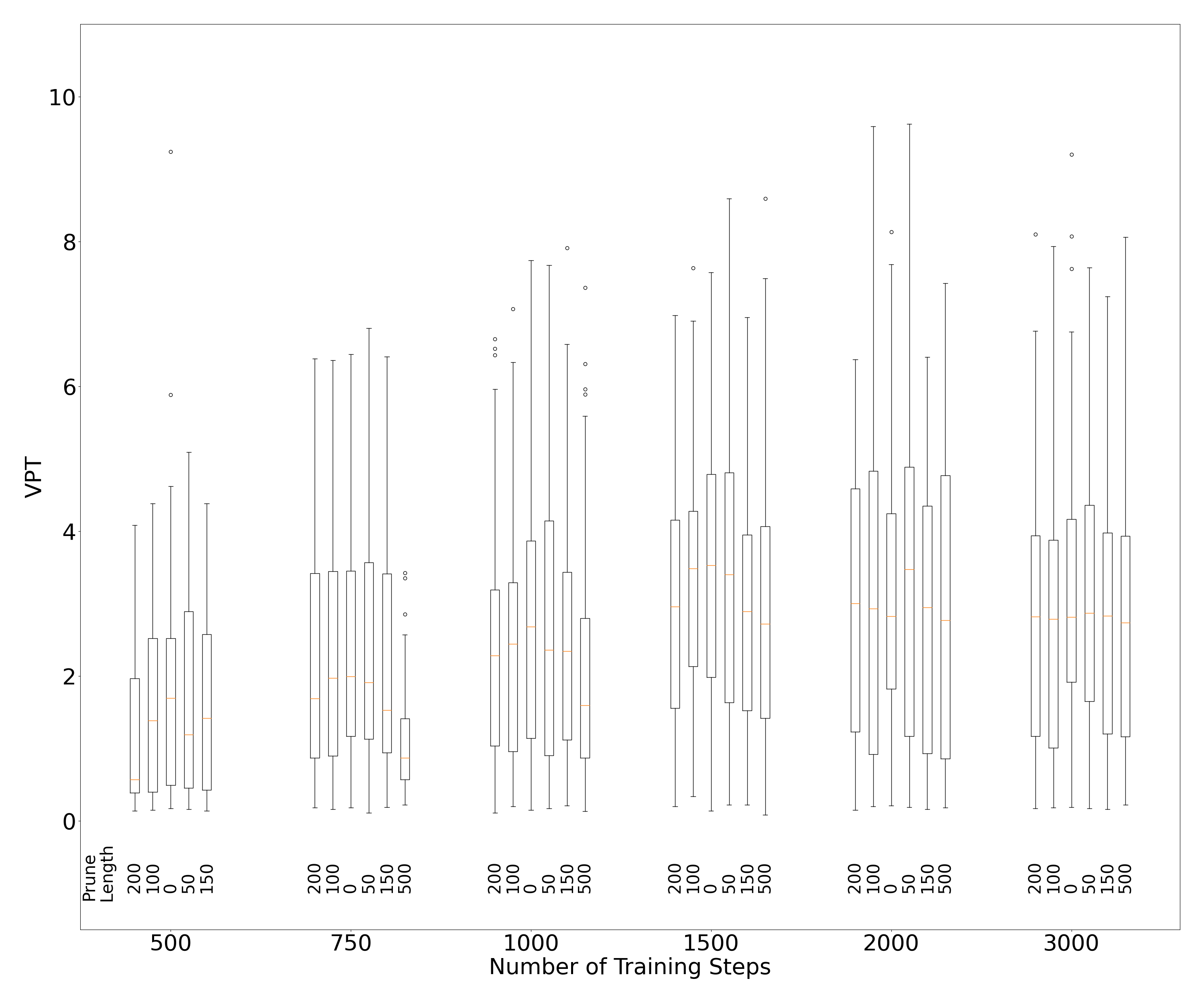}
    \caption{Box plot with the same box structure as in Fig.~\ref{fig:box_fixed_layers} depicting VPT values for statistical data for different training and pruning length. }
    \label{fig:train_prune}
\end{figure}

Additionally, non-negligible effect can be observed by slecting different training lengths\footnote{In these experiments we have the comfort to manipulate with that hyperparameter, however, in the real world scenario one may be restricted by the available data.} and pruning lengths. The results from that analysis are collected in Fig.~\ref{fig:train_prune}. 

The results of this analysis are short of being conclusive, with slight preference towards longer training sets. The pruning length, however, does not play as significant role as in the case of classical reservoir computing, as long as the ration between pruning and training length is reasonably small (based on our observations, we suggest limiting prune/train length to $<$0.25). 

\begin{figure*}[!htbp]
    \centering
    \includegraphics[width = 0.45\textwidth]{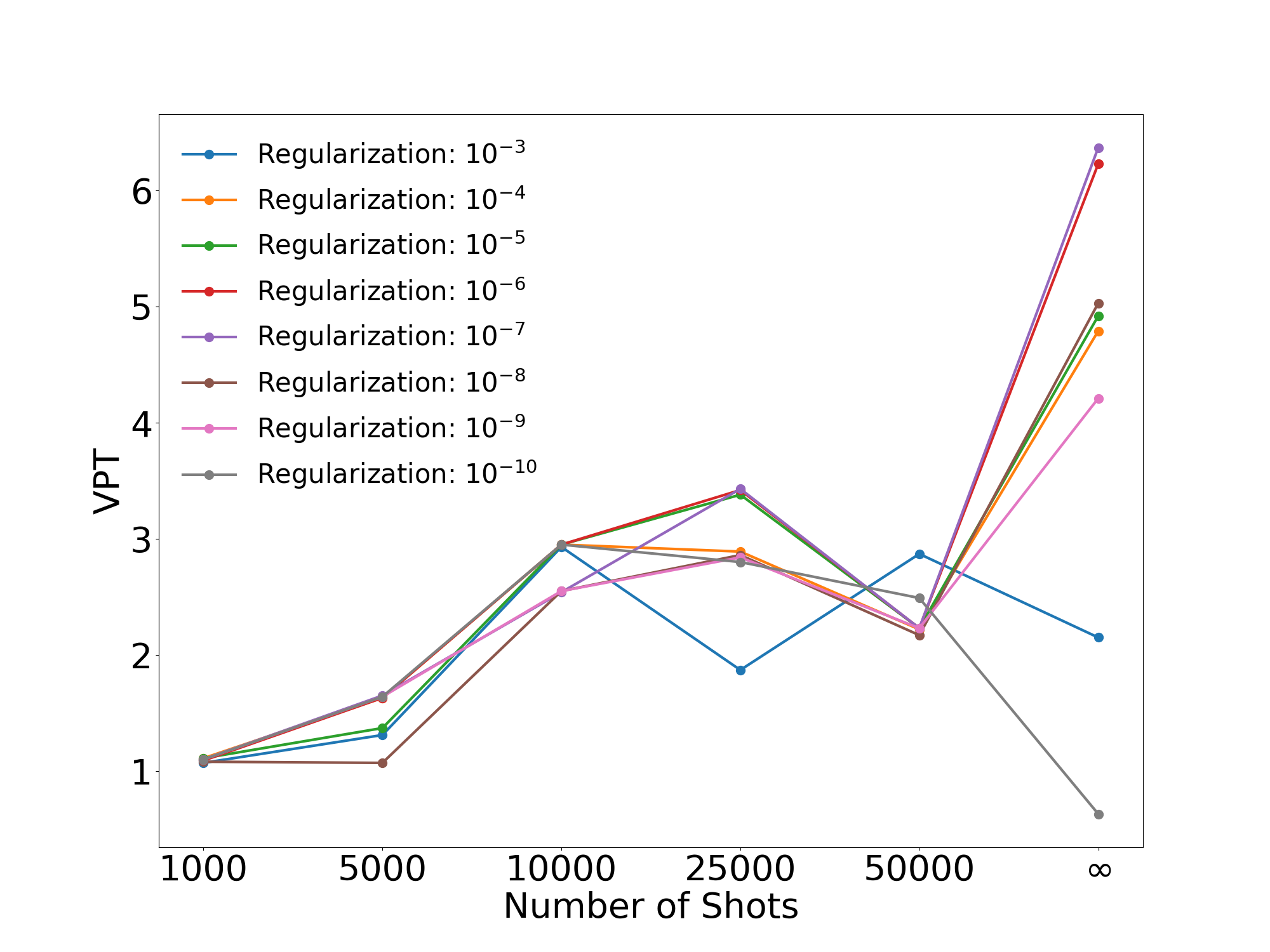}
    \includegraphics[width = 0.45\textwidth]{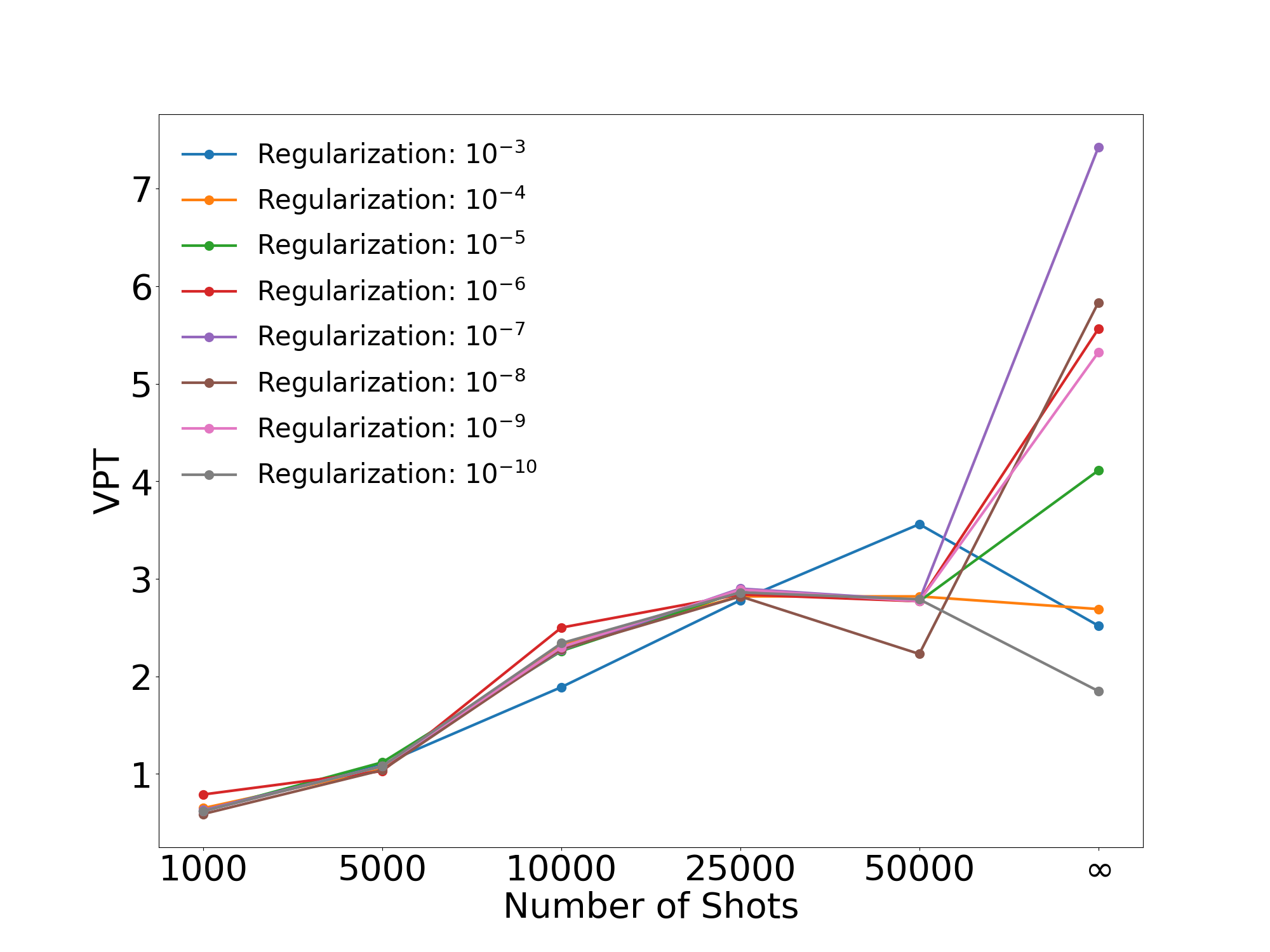}
    \caption{Predictive power of the setups expressed by the VPT values as a function of number of shots taken to estimate expectation values for setups with different regularization value. Plots for leak rates: (left) $\alpha = 0.6$ and (right) $\alpha=0.7$. }
    \label{fig:reg_shots}
\end{figure*}

In Fig.~\ref{fig:reg_shots} we investigate how different number of shots can affect performance for different values of regularization for setups with $\alpha = 0.6$ (left panel) and $\alpha = 0.7$ (right panel). 

We see that increasing number of shots towards the exact values of expectation values (i.e. shots$\to\infty$) we improve the VPT values across all setups, except the smallest and largest tested regularization values. Though the trend is non-monotonic due to statistical variability in collecting samples. 


\bibliography{refs}

\end{document}